\documentclass[traditabstract]{aa}
\usepackage{psfig}
\usepackage{graphicx}

\begin{document}

\title{Accurate classification of 17 AGNs detected with
Swift/BAT\thanks{
Based on observations obtained from the following observatories: Telescopio Nazionale Galileo at the Observatorio del
Roque de los Muchachos of the Instituto de Astrof\'isica de Canarias (Canary Islands, Spain); 
Astronomical Observatory of Bologna in Loiano (Italy); Astronomical Observatory of Asiago (Italy);
Cerro Tololo Interamerican Observatory (Chile); Observatorio Astron\'omico Nacional (San Pedro
M\'artir, Mexico).}}

%\subtitle{}

\author{P. Parisi\inst{1,} \inst{2}, N. Masetti\inst{1}, E. Jim\'enez-Bail\'on\inst{3}, V. Chavushyan\inst{4}, A. Malizia\inst{1}, R. Landi\inst{1}, M. Molina\inst{5}, M. Fiocchi\inst{6}, E. Palazzi\inst{1},
L. Bassani\inst{1}, A. Bazzano\inst{6}, A.J. Bird\inst{7}, A.J. Dean\inst{7}, G. Galaz\inst{8}, E. Mason\inst{9}, D. Minniti\inst{8,} \inst{10}, L. Morelli\inst{11}, J.B. Stephen\inst{1}, and
P. Ubertini\inst{6}
}
\institute{
INAF -- Istituto di Astrofisica Spaziale e Fisica Cosmica di 
Bologna, Via Gobetti 101, I-40129 Bologna, Italy
\and
Dipartimento di Astronomia, Universit\`a di Bologna, Via Ranzani 1,
I-40129 Bologna, Italy
\and 
Instituto de Astronom\'{\i}a, Universidad Nacional Aut\'onoma de M\'exico,
Apartado Postal 70-264, 04510 M\'exico D.F., M\'exico
\and
Instituto Nacional de Astrof\'{i}sica, \'Optica y Electr\'onica,
Apartado Postal 51-216, 72000 Puebla, M\'exico
\and 
INAF-- Istituto di Astrofisica Spaziale e Fisica Cosmica di 
Milano, Via Bassini 15, 20122 Milan, Italy
\and
Istituto di AstroÞsica Spaziale e Fisica Cosmica di Roma (INAF), Via Fosso del Cavaliere 100, Roma I-00133, Italy
\and
School of Physics \& Astronomy, University of Southampton, Southampton, Hampshire, SO171BJ, United Kingdom
\and
Departamento de Astronom\'{i}a y Astrof\'{i}sica, Pontificia Universidad 
Cat\'olica de Chile, Casilla 306, Santiago 22, Chile                     
\and
European Southern Observatory, Alonso de Cordova 3107, Vitacura,
Santiago, Chile
\and
Specola Vaticana, V-00120 Citt\`a del Vaticano
\and
Dipartimento di Astronomia, Universit\`a di Padova,
Vicolo dell'Osservatorio 3, I-35122 Padua, Italy
}

\offprints{P. Parisi (\texttt{parisi@iasfbo.inaf.it)}}
\date{Received 20 July 2009; accepted 12 September 2009}

\abstract{Through an optical campaign performed at 5 telescopes located in the northern and the southern hemispheres, plus 
archival data from two on line sky surveys, we have obtained optical spectroscopy for 17 counterparts of suspected or poorly studied 
hard X--ray emitting active galactic nuclei (AGNs) detected with {\it Swift}/BAT in order to determine or better classify their nature. 
We find that 7 sources of our sample are Type 1 AGNs, 9 are Type 2 AGNs, and 1 object is an X-ray bright optically normal galaxy; 
the redshifts of these objects lie in a range between 0.012 and 0.286. For all these sources, X-ray data analysis was also performed 
to estimate their absorption column and to search 
for possible Compton thick candidates. Among our type 2 objects, we did not find any clear Compton thick AGN, but at least 6 out of 9 of them are 
highly absorbed (N$_{H}$$>$ 10$^{23}$ cm$^{-2}$), while one does not require intrinsic absorption; i.e., it appears to be a naked 
Seyfert 2 galaxy.

}%{}{}{}{}


\titlerunning{Accurate classification of 17 AGNs detected with Swift/BAT}
%\titlerunning{A multi-observatory identification campaign on 
%{\it INTEGRAL} sources}
\authorrunning{P. Parisi et al.}

\maketitle

\section{Introduction}
The {\it Swift} mission was designed to study cosmic gamma-ray bursts (GRBs) in a multiwavelength context (Gehrels et al. 2004).
The aims of this mission are to determine of the origin of these phenomena and to search for new types of GRBs, including their interaction with the surrounding 
medium and their use as probes to study the universe at z $>$ 5. 

{\it Swift}, with its unique repointing capabilities, is able to study and monitor other types of X-ray emitting objects.
Through its payload, consisting of three instruments, i.e. the burst alert telescope (BAT; Barthelmy, 2004), the X-Ray telescope (XRT; Burrows et al. 2004) and the ultraviolet/optical telescope (UVOT; Roming et al. 2004), {\it Swift} can detect and follow up X-ray emitting objects up over a wide range of wavelengths.

In particular, BAT is a coded mask instrument operating in the energy range 14--195 keV, with a field of view of 1.4 sr, 
and is able to provide a source position determination with an uncertainty of $1^{\prime}-4^{\prime}$  (Gehrels et al. 2004) depending on source intensity.
Its sensitivity is estimated at $\sim$1 mCrab at high Galactic latitudes and $\sim$3 mCrab for strong sources on the Galactic plane.

As said before, this instrument is not only able to detect new GRBs, but also to perform a highly sensitive 
hard X-ray survey of the sky (e.g. Cusumano et al. 2009). Indeed, the BAT surveys of Tueller et al. (2008, 2009) allow, in particular, study of the extragalactic X-ray sky, and the observation of absorbed AGNs over a range of energies not affected by absorption due to intervening material.

Observations performed below 10 keV with other satellites, such as {\it ASCA}, {\it BeppoSAX} (Matt et al. 2000, Ueda et al. 1999), {\it Chandra}, {\it XMM}, and {\it Suzaku} (Ueda et al. 2007, Guainazzi et al. 2005), have revealed a population of absorbed AGNs with a hydrogen column 
density along the line-of-sight higher than $10^{23}$ cm$^{-2}$, which obscures the nuclei at optical and soft (0.2-10 keV) X-ray bands.
Quantifying the number of these particular objects, especially at low redshifts, is very important if one wants to understand the accretion \mbox{mechanisms} at work in AGNs and how the 
absorbed AGNs contribute to the cosmic X--ray background (Comastri et al. 2004).

Now a number of surveys at energies higher than 10 keV are available to study this class of objects. The surveys performed by IBIS (Ubertini et al. 2003) on board
{\it INTEGRAL} (Winkler et al. 2003), together with those of BAT, provide the best sample of objects selected in the soft gamma-ray band to date (Bird et al. 2007, Krivonos et al. 2007).
IBIS and BAT work in similar spectral bands, but concentrate on different parts of the sky: IBIS maps mainly the Galactic plane, while BAT focuses on observations at high Galactic latitudes.\\

The nature of AGNs detected in these surveys is however often not confirmed and sometimes just assumed on the basis of their X-ray spectrum; therefore an optical follow-up of these sources is required.
The optical spectra are not only crucial for an accurate classification, but can provide fundamental parameters which together with softness flux ratio (as in Malizia et al. 2007), can give us information 
about their possible Compton thick nature (an AGN is defined as `Compton thick' when the column density along its line of sight is equal to or greater than the inverse of the Thomson cross-section, i.e. $N_H \geq 1.5\times10^{24}$ cm$^{-2}$).

In this work we have selected from the {\it Swift}/BAT AGN surveys of Tueller et al. (2009), Ajello et al. (2008) and Winter et al. (2008) those objects (17 in total) either without optical identification,
or not well studied or without published optical spectra.
Following the method applied by Masetti et al. (2004, 2006a,b, 2008, 2009) for the optical spectroscopic follow-up of unidentified {\it INTEGRAL} sources, we determine the nature
of the 17 selected objects which are listed in Table \ref{log}.

We show here the spectroscopic results obtained on this sample at various telescopes, thanks to an observational multisite campaign carried out in Europe, Central
and South America plus the use of archival spectra available online. In Sect. 2 we give a description of the observations and the employed telescopes, with information on the data reduction method. 
Sect. 3 reports the X-ray data analysis for all of our sources. Sect. 4 reports and discusses the results for each individual source together with general characteristics (central black hole masses and Compton thickness). In Sect. 5 comments and conclusions are given. 
\section{Optical spectroscopy}
The following telescopes were used for the optical spectroscopic study presented here:

\begin{itemize}
\item the 1.5m telescope at the Cerro Tololo Interamerican Observatory (CTIO), Chile;
\item the 1.52m ``Cassini'' telescope of the Astronomical Observatory of 
Bologna, in Loiano, Italy; 
\item the 1.8m ``Copernicus'' telescope at the Astrophysical Observatory 
of Asiago, in Asiago, Italy;
\item the 2.1m telescope of the Observatorio Astr\'onomico Nacional in San Pedro Martir, Mexico;
\item the 3.58m telescope ``Telescopio Nazionale Galileo'' (TNG) at the Observatorio of the la Roque de Los Muchachos (Canary Islands, Spain); 
\end{itemize}
The data reduction was performed with the standard procedure (optimal extraction; Horne 1986) using IRAF\footnote{
IRAF is the Image Reduction and Analysis 
Facility made available to the astronomical community by the National 
Optical Astronomy Observatories, which are operated by AURA, Inc., under 
contract with the U.S. National Science Foundation. It is available at 
{\tt http://iraf.noao.edu/}}.
Calibration frames (flat fields and bias) were taken on the day preceeding or following 
the observing night. The wavelength calibration was obtained using lamp spectra 
acquired soon after each on-target spectroscopic acquisition; the uncertainty on the 
calibration was $\sim$0.5~\AA~for all cases; this was checked using the positions of 
background night sky lines. Flux calibration was performed using 
catalogued spectrophotometric standards.
Objects with more than one observation had their spectra stacked 
together to increase the signal-to-noise ratio.

Further spectra were retrieved from two different astronomical archives:
the Sloan Digitized Sky Survey\footnote{{\tt 
http://www.sdss.org}} (SDSS, Adelman-McCarthy et al. 2005) archive, and 
the Six-degree Field Galaxy Survey\footnote{{\tt 
http://www.aao.gov.au/local/www/6df/}} (6dFGS) archive (Jones et al. 
2004). As the 6dFGS archive provides spectra which are not calibrated in 
flux, we used the optical photometric information in Jones et al. (2005) 
and Doyle et al. (2005) to calibrate the 6dFGS data presented here.
The only exception was source Swift J0342.0$-$2115, for which we used the optical photometric 
information available in SIMBAD\footnote{available at {\tt http://simbad.u-strasbg.fr/simbad/}}
because its {\it R} magnitude, essential for the spectrum flux calibration, was not reported in Jones et al. (2005).

The identification and classification approach we adopt in the analysis of the optical spectra is the following:
for the emission-line AGN classification, we used the criteria 
of Veilleux \& Osterbrock (1987) and the line ratio diagnostics 
of Ho et al. (1993, 1997) and of Kauffmann et al. (2003); for 
the subclass assignation to Seyfert 1 galaxies, we used the 
\mbox{H$_\beta$/[O {\sc iii}]$\lambda$5007} line flux ratio criterion as in
Winkler et al. (1992).

When possible, in order to provide an estimate of the local absorption in the Seyfert 2 galaxies of our sample and an assessment
of the Compton nature of these AGNs, we first dereddened the  H$_\alpha$ and H$_\beta$ line 
fluxes by applying a correction for the Galactic absorption along the line of sight of the source. This was possible 
by using the estimate of the Galactic color excess $E(B-V)_{\rm Gal}$ given by Schlegel et al. (1998), and
the Galactic extinction law by Cardelli et al. (1989).
Then, we estimated the color excess  \mbox{$E(B-V)_{\rm AGN}$} local to the AGN host from the comparison between the intrinsic 
line ratio and that corrected for Galactic reddening, using again the extinction law of Cardelli et al. (1989) and 
assuming an intrinsic H$_\alpha$/H$_\beta$ line ratio of 2.86 (Osterbrock 1989). 

The spectra of these objects are not corrected for starlight contamination 
(see, e.g., Ho et al. 1993, 1997), because of their limited S/N and the 
spectral resolution. This however does not affect our results and conclusions.
In this work we consider a \mbox{cosmology} with $H_{\rm 0}$ = 65
km s$^{-1}$ Mpc$^{-1}$, $\Omega_{\Lambda}$ = 0.7 and $\Omega_{\rm m}$ =
0.3; the luminosity distances of the objects reported in 
this paper are computed from these parameters using the Cosmology
Calculator of Wright (2006).

\begin{figure*}%[th!]
%\begin{center}
\hspace{-.1cm}
\centering{\mbox{\psfig{file=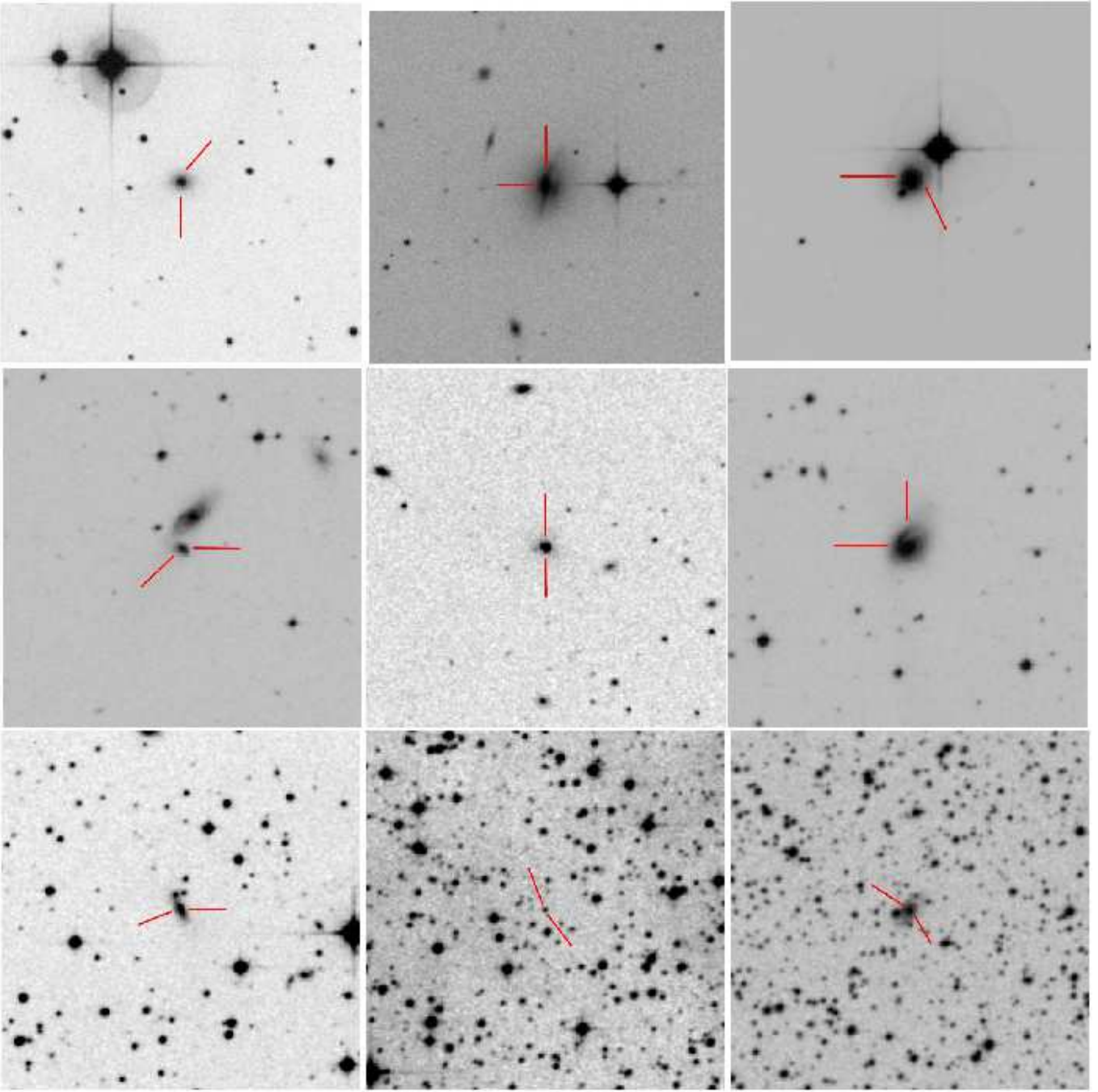,width=18cm}}}

\caption{From left to  right and top to bottom:  optical images of the
fields  of Swift J0059.4+3150, Swift J0134.1$-$3625, Swift J0342.0$-2115$, Swift J0350.1$-$5019, Swift J0505.7$-$2348, 
Swift J0501.9$-$3239, Swift J0640.1$-$4328,
Swift J0727.5$-$2406 and Swift J0739.6$-$3144.
The optical counterparts  of the {\it Swift} sources are indicated with tick marks. Field sizes are
5$'$$\times$5$'$ and are extracted  from the DSS-II-Red survey. In all
cases, North is up and East to the left.}\label{fields}
%\end{center}
\end{figure*}

\begin{figure*}%[th!]
%\begin{center}
%\centering{\mbox{\psfig{file=swiftj0854.7.ps,width=5.9cm}}}

%\vspace{-.3cm}
\centering{\mbox{\psfig{file=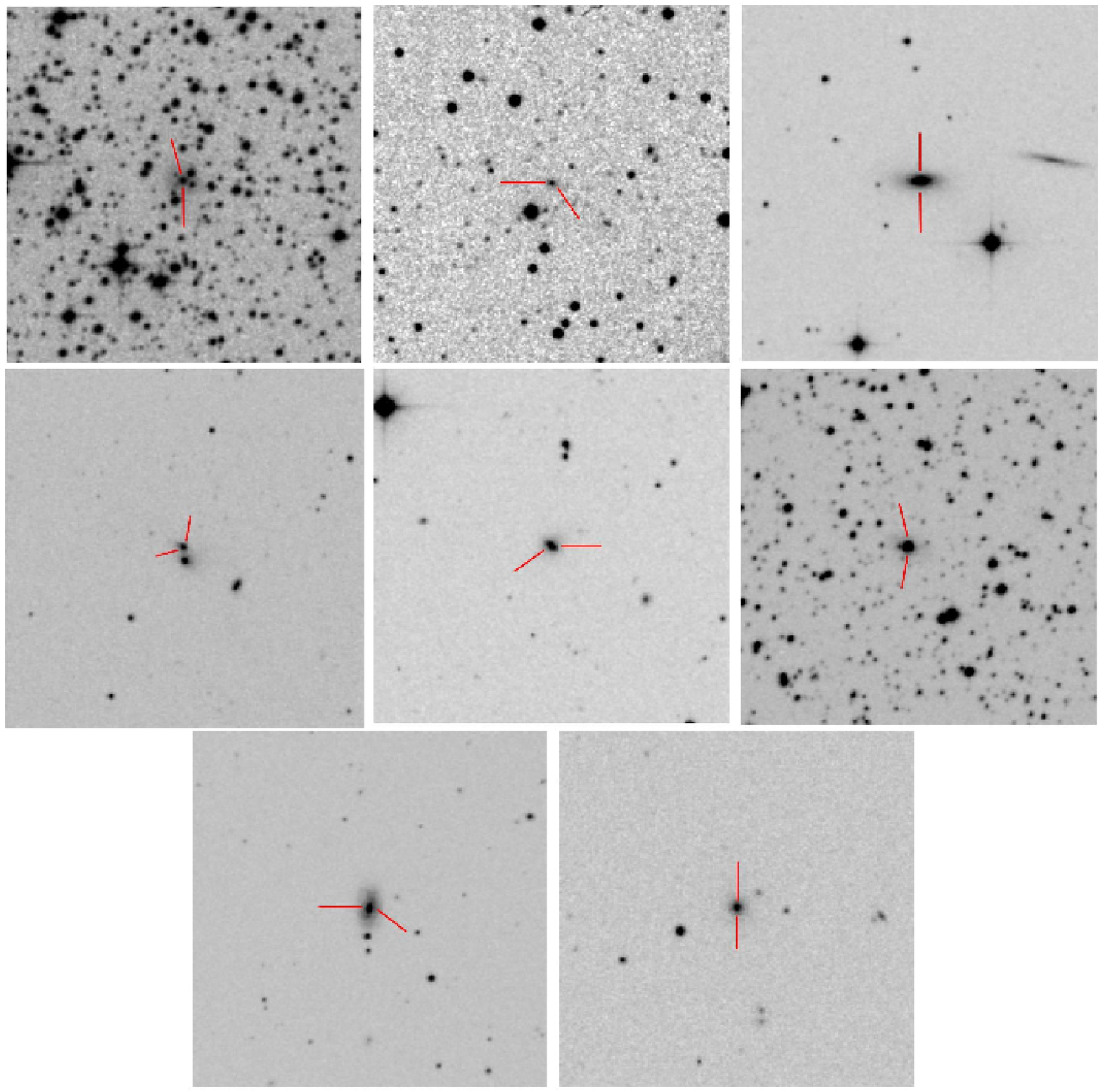,width=18cm}}}

%\vspace{-.3cm}
%\centering{\mbox{\psfig{file=swij2319.4.ps,width=5.9cm}}}
%\hspace{.5cm}
%\parbox{11.5cm}{
%\vspace{-.5cm}
\caption{As Fig.~\ref{fields}, for the fields of Swift J0743.0$-$2543, Swift J0811.5+0937, Swift J0902.0+6002.Swift J0904.3+5538, Swift J 0911.2+4533, Swift J0917.2$-$6221, Swift J0923.7+2255 and Swift J1049.4+2258. 
The optical counterparts  of the {\it Swift} sources are indicated with tick marks. Field sizes are
5$'$$\times$5$'$ and are extracted  from the DSS-II-Red survey. In all cases, North is up and East to the left.}\label{fields2}
%\end{center}
\end{figure*}

In Figures \ref{fields} and \ref{fields2} we show the optical finding charts of the 17 sources presented in this work.
The corresponding optical counterparts are indicated with tick marks.
In Table~\ref{log} the detailed log of observations is reported.

We list in column 1 the name of the observed {\it Swift} sources. 
In columns 2 and 3 we report the coordinates of the objects obtained from the 2MASS catalog\footnote{available at {\tt http://www.ipac.caltech.edu/2mass/}} (Skrutskie et al. 2006). 
In column 4 we list the telescope and the instrument used for the source observation. The characteristics of each 
spectrograph are given in columns 5 and 6. Column 7 provides the 
observation date and the UT time at mid-exposure, while column 8 reports 
the exposure times and the number of spectral pointings.
\begin{table*}[th!]
\caption[]{Log of the spectroscopic observations presented in this paper
(see text for details). Source coordinates
are extracted from the 2MASS catalog and have an accuracy better than 0$\farcs$1.} \label{log}
\scriptsize
%\hspace{-.3cm}
%\vspace{-1.3cm}
\begin{center}
\begin{tabular}{llllcccr}
\noalign{\smallskip}
\hline
\hline
\noalign{\smallskip}
\multicolumn{1}{c}{{\it (1)}} & \multicolumn{1}{c}{{\it (2)}} & \multicolumn{1}{c}{{\it (3)}} & \multicolumn{1}{c}{{\it (4)}} & 
{\it (5)} & {\it (6)} & {\it (7)} & \multicolumn{1}{c}{{\it (8)}} \\
\multicolumn{1}{c}{Object} & \multicolumn{1}{c}{RA} & \multicolumn{1}{c}{Dec} & 
\multicolumn{1}{c}{Telescope+instrument} & $\lambda$ range & Disp. & \multicolumn{1}{c}{UT Date \& Time}  & Exposure \\
& \multicolumn{1}{c}{(J2000)} & \multicolumn{1}{c}{(J2000)} & & (\AA) & (\AA/pix) & 
\multicolumn{1}{c}{at mid-exposure} & time (s)  \\

\noalign{\smallskip}
\hline
\noalign{\smallskip}

Swift J0059.4+3150         &00 59 53.28& +31 49 37.3   & SPM 2.1m+B\&C Spc. &3450-7650& 4.0 & 30 Jun 2008, 10:50 & 1800  \\
Swift J0134.1$-$3625     &01 33 57.75& $-$36 29 35.8   & CTIO 1.5m+RC Spc.   & 3300-10500 & 5.7 & 02 Jul 2008, 09:02 & 1200 \\
Swift J0342.0$-$2115      & 03 42 03.71 &$-$21 14 39.6&  AAT+6dF    & 3900-7600 & 1.6& 30 Nov 2003, 13:34 & 1200+600   \\
Swift J0350.1$-$5019      & 03 50 23.78 &$-$50 18 35.5  &  CTIO 1.5m+RC Spc.   & 3300-10500 & 5.7 &04 Mar 2008, 01:23 & 2$\times$1800  \\
Swift J0505.7$-$2348      & 05 05 45.75 &$-$23 51 14.0& SPM 2.1m+B\&C Spc. &3450-7650& 4.0 & 03 Feb 2008, 04:04 & 2$\times$1800       \\
Swift J0501.9$-$3239$^*$ & 05 19 35.81&$-$32 39 28.0 & AAT+6dF    & 3900-7600 & 1.6 & 30 Jen 2003, 10:36 & 1200+600   \\
Swift J0640.1$-$4328       & 06 40 37.99 &$-$43 21 21.1  & CTIO 1.5m+RC Spc.   & 3300-10500 & 5.7& 05 Mar 2008, 1:26 & 2$\times$1800  \\
%Swift J0641.3+3257          & 06 41 23.06 &+32 55 38.4  & Cassini+BFOSC & 3500-8000 & 4.0 &22 Dec 2007, 23:53& 2$\times$1200 \\
Swift J0727.5$-$2406       & 07 27 21.05& $-$24 06 32.3  & TNG+DOLoRes   & 3800-8000 & 2.5  & 07 Feb 2008,  23:35 & 2$\times$1200  \\
Swift J0739.6$-$3144       & 07 39 44.69 &$-$31 43 02.5 & SPM 2.1m+B\&C Spc. &3450-7650& 4.0 & 03 Feb 2008, 07:15 & 2$\times$1800  \\
Swift J0743.0$-$2543      & 07 43 14.72 &$-$25 45 50.1 & CTIO 1.5m+RC Spc.   & 3300-10500 & 5.7 & 04 Mar 2008, 03:56 & 2$\times$1200  \\
Swift J0811.5+0937         & 08 11 30.83 &+09 33 50.9 & TNG+DOLoRes   & 3800-8000 & 2.5 & 08 Feb 2008, 00:51 & 2$\times$1800  \\
%Swift J0854.7+1502         & 08 54 29.70 &+15 01 36.2 & SPM 2.1m+B\&C Spc. &3450-7650& 4.0 & 02 Feb 2008, 09:17 & 2$\times$1800  \\
Swift J0902.0+6007          & 09 01 58.37 &+60 09 06.2& Copernicus+AFOSC & 4000-8000 & 4.2  & 22 Feb 2008, 21:07 & 1200+600       \\
Swift J0904.3+5538        &09 04 36.97 &+55 36 02.6   & SDSS+CCD Spc. & 3800-9200 & 1.0 & 29 Dec 2000, 10:48 & 9000   \\
Swift J0911.2+4533        &09 11 30.00& +45 28 06.0 & SDSS+CCD Spc.&3800-9200&1.0&  07 Feb 2002, 08:24  & 4803 \\
Swift J0917.2$-$6221     & 09 16 09.37& $-$62 19 29.6& CTIO 1.5m+RC Spc.   & 3300-10500 & 5.7  & 04 Mar 2008, 06:02 & 600  \\
Swift J0923.7+2255          &09 23 43.01 &+22 54 32.4   &  SPM 2.1m+B\&C Spc. &3450-7650& 4.0 & 02 Feb 2008, 11:23 & 1800           \\
Swift J1049.4+2258         & 10 49 30.89& +22 57 52.4   & Cassini+BFOSC & 3500-8000 & 4.0 & 08 May 2008, 21:17 & 2$\times$1200  \\
%Swift J2319.4+2619       & 23:19:30.43 & +26:15:19.1 & SPM 2.1m+B\&C Spec. &3450-7650& 4.0 & 28 Jun 2008, 09:27 & 2$\times$1800  \\
\noalign{\smallskip}
\hline
\noalign{\smallskip} 
\multicolumn{8}{l}{$^*$ The name of this source reported in Tueller et al. (2009) does not correspond to the actual coordinates.} \\ 
%\multicolumn{8}{l}{} \\
%\multicolumn{8}{l}{}\\
\noalign{\smallskip}
\noalign{\smallskip}
\end{tabular}
\end{center}
\end{table*}

\section{X-ray data analysis}
Here we report the X-ray data analysis of the 17 sources identified in this work. 
It is worth noting that for most of the sources the X-ray spectral parameters ($\Gamma$, N$_H$ and possibly the 2-10 keV flux) 
were already published by various authors; here for consistency we decided to re-analyze the X-ray data in order to always apply
the same simple model (an absorbed power law) and to properly use the values obtained in the diagnostic diagram of Malizia et al.(2007 see Section 4.19). 
For 13 out of 17 objects, data acquired with the XRT were analyzed, while for the remain archival XMM-\emph{Newton} and \emph{Chandra}
observations have been considered.

For Swift J0739.6-3144 XMM-\emph{Newton} EPIC-pn (Struder at al. 2001) data were analyzed (see subsection 4.9) 
while for Swift J0743.0-2543 an XMM-\emph{Newton} slew pointing is available and therefore only the 2-10 keV flux 
value can be used.
The two remaining sources, Swift J0727.5-2406 and Swift J0811.5+0937, were observed with the \textit{Chandra X-ray Observatory} 
for 5126 s (start time 07:49:36 UT of 2007-12-20) and 5139 s (start time 15:04:24 UT of 2007-12-27), respectively, but since both the sources are very dim,
we were only able to estimate their 2-10 keV fluxes using the {\it eff2evt} tool of the CIAO {\it Chandra} analysis software.

The \textit{Chandra} data reduction was performed using CIAO v4.1 software with the
calibration database CALDB v4.1, provided by the \textit{Chandra}
X-ray Center and following the science threads listed on the CIAO
website\footnote{Available at http://cxc.harvard.edu/ciao/.}.

XRT data reduction was performed using the XRTDAS standard data pipeline package ({\sc xrtpipeline} v.
0.12.1) in order to produce screened event files. All data were extracted only
in the photon counting (PC) mode (Hill et al. 2004), adopting the standard grade filtering (0--12 for
PC) according to the XRT nomenclature. 
For observations in which the XRT count rate was high enough to produce event pile-up, we extracted 
the events in an annulus centered on the source, determining the size of the inner circle according 
to the procedure described in Romano et al. (2006). In the other cases, events for spectral analysis 
were extracted within a circular region of radius 20$^{\prime \prime}$, centered on the source 
position, which encloses about 90\% of the PSF at 1.5 keV (see Moretti et al. 2004).
The background was taken from various source-free regions close to the X-ray source of interest, using 
circular regions with different radii in order to ensure an evenly sampled background. In all cases, 
the spectra were extracted from the corresponding event files using the {\sc XSELECT} software and 
binned using {\sc grppha} in an appropriate way, so that the $\chi^{2}$ statistic could be applied. We 
used version v.011 of the response matrices and create individual ancillary response files
using {\sc xrtmkarf v. 0.5.6}.

XMM-\emph{Newton} EPIC-pn (Struder et al. 2001) data reduction was performed
using the \emph{XMM} standard analysis software (SAS) version 8.0
employing the latest available calibration files. Only patterns
corresponding to single and double events (PATTERN$\leq$4) were taken
into account; the standard selection filter FLAG=0 was applied. The
observations were filtered for periods of high background. 
Source counts were extracted from a circular region of radius 25$^{\prime\prime}$ centred on the source in
order to exclude the extended emission associated with the galaxy;
background spectra were extracted from circular regions close to the
source, or from source-free regions of 80$^{\prime\prime}$ radius. The
ancillary response matrices and the detector response matrices 
were generated using the \emph{XMM} SAS tasks \emph{arfgen} and
\emph{rmfgen}; spectral channels were rebinned in order to achieve a
minimum of 20 counts for each bin. 

Spectral analyses were performed using {\sc XSPEC} version 12.5.0 and all errors are quoted 
at 90\% confidence level for one parameter of interest ($\Delta \chi^{2}$=2.71).
For sources with more than one pointing, we performed the spectral analysis of the combined spectra in order to 
improve the statistical quality of the data.
Due to the low statistics available, we identified for each source the best energy range for the spectral analysis and 
we employed a simple power law (often fixing the photon index to a canonical value of 1.8),
absorbed by both the Galactic (Dickey \& Lockman 1990) and an intrinsic neutral hydrogen column density as our baseline model.
\begin{table*}[th!]
\begin{center}
\caption[]{Main results obtained from the analysis of the optical spectra of the 17 AGNs of the present 
sample of {\it Swift} sources.}\label{res}
\scriptsize
\begin{tabular}{lcccccrccl}
\noalign{\smallskip}
\hline
\hline
\noalign{\smallskip}
\multicolumn{1}{c}{Object} & $F_{\rm H_\alpha}$ & $F_{\rm H_\beta}$ &
$F_{\rm [OIII]}$ & Class & $z$ & \multicolumn{1}{c}{$D_L$} & \multicolumn{2}{c}{$E(B-V)$}&\multicolumn{1}{c}{ L$_{Radio}$} \\
\cline{8-9}
\noalign{\smallskip}
& & & & & & (Mpc) & Gal. & AGN&\multicolumn{1}{c}{(erg s$^{-1}$) } \\
\noalign{\smallskip}
\hline
\noalign{\smallskip}

Swift J0059.4+3150 & * & 87$\pm$5& 37$\pm$2 & Sy1.2 & 0.015 & 84.2 & 0.061& -- &\multicolumn{1}{c}{--}\\
& * & [105$\pm$6] & [45$\pm$3] & & & & & & \\
& & & & & & & & &\\
& & & & & & & & &\\

Swift J0134.1$-$3625 & 9.4$\pm$0.3 & in abs.& 5.3$\pm$2.2 & Sy2 & 0.029 & 136.7 & 0.020 &--& \multicolumn{1}{c}{--}\\
& [9.9$\pm$0.4] & [in abs.] & [5.6$\pm$2.3] & & & & & &\\
% & & & & & & & & 1100 (20--100) \\

& & & & & & & & &\\

Swift J0342.0$-$2115 & * & 2000$\pm$400 & 400$\pm$80& Sy1 & 0.0139 & 67.6 & 0.032 &--&\multicolumn{1}{c}{--} \\
& * & [2700$\pm$600] & [460$\pm$90] & & & & & &\\

& & & & & & & & &\\

Swift J0350.1$-$5019 & 38$\pm$3 & 6.9$\pm$0.9 & 11$\pm$1 & Sy2 & 0.035 & 165.8 & 0.015 & 0.216&\multicolumn{1}{c}{--}  \\
& [39$\pm$3] & [7.2$\pm$0.9] & [11$\pm$1] & & & & & &\\

& & & & & & & & & \\

Swift J0505.7$-$2348 & 38$\pm$1 & 9$\pm$1 & 48$\pm$1 & Sy2 & 0.036 & 172.1 & 0.040 & 0.144&3.1$\times$10$^{38}$ (1.4 GHz)  \\
& [41$\pm$1] & [10$\pm$1] & [55$\pm$1] & & & & & & \\

& & & & & & & & &\\

Swift J0501.9$-$3239 & 180$\pm$20 & 52$\pm$11 & 320$\pm$30 & Sy2 & 0.0126 & 58.9 & 0.017&0.036&7.0$\times$10$^{37}$ (843 MHz)  \\
& [190$\pm$20] & [59$\pm$13] & [350$\pm$30] & & & & & &\\

& & & & & & & & &\\

Swift J0640.1$-$4328 & 3.1$\pm$0.5 & 0.7$\pm$0.3 &2.3$\pm$0.4 & Sy2& 0.061 &293.9 & 0.084&--&8.1$\times$10$^{39}$ (843 MHz)  \\
& [4.1$\pm$0.7] & [3.6$\pm$1.7] & [3.0$\pm$0.6] & & & & & &\\

& & & & & & & & &\\
% Swift J0641.3+3257& in abs.& in abs. & $<$3.7 & XBONG & 0.016 & 74.7 & 0.149 &-- & 22.53 (2--10) \\
% & [in abs.]& [in abs.] & [$<$8.5] & & & & & & 3.67 (14--195) \\
%& & & & & & & & & \\
Swift J0727.5$-$2406 & *  & 2.1$\pm$0.7 & 21$\pm$1 & Sy1.9& 0.123& 619.0 & 1.047&--&1.9$\times$10$^{40}$ (1.4 GHz)  \\
& * & [47$\pm$7] & [385$\pm$17] & & & & & &\\

& & & & & & & & &\\

Swift J0739.6$-$3144& 36$\pm$2 & 2.3$\pm$0.5 & 32$\pm$1 & Sy2 & 0.026 & 122.3 & 0.613 & 0.360&2.5$\times$10$^{38}$ (1.4 GHz) \\
& 142$\pm$7 & [18$\pm$4] & [216$\pm$5] & & & & & &6.4$\times$10$^{38}$ (843 MHz)\\

& & & & & & & & &\\

Swift J0743.0$-$2543 & * & 11$\pm$2 & 4.5$\pm$0.7 & Sy1.2  & 0.023 & 108.0 & 0.678 & --&1.2$\times$10$^{37}$ (1.4 GHz) \\
& *& [146$\pm$20] & [38$\pm$6] & & & & & &\\

& & & & & & & &  &\\

Swift J0811.5+0937  & -- & in abs.&--& XBONG & 0.286 & 1582.6 & 0.026& --&1.8$\times$10$^{40}$ (1.4 GHz)  \\
& & [in abs.] & & & & & & &\\

& & & & & & & & &\\

%Swift J0854.7+1502  & 7.1$\pm$1.2& 0.9$\pm$0.1 & 4.9$\pm$0.3 & Sy2  & 0.071 & 345.0 & 0.040 &0.306 & 24.4 (14--195) \\
% & [7.2$\pm$1.2]& [1.1$\pm$0.2] & [5.5$\pm$0.3] & & & && &  \\
% & & & & & & & & & \\
Swift J0902.0+6007  & *& 52$\pm$5 & 98$\pm$5& Sy2  & 0.012 & 55.9 & 0.043 &0.198&1.6$\times$10$^{38}$ (1.4 GHz)  \\
& *& [58$\pm$6] & [114$\pm$6] & & & & & &\\

& & & & & & & & &\\

Swift J0904.3+5538& *& 30.9$\pm$7.9 &23.1 $\pm$0.8 & Sy1.5  & 0.0374 & 177.4 & 0.021& --&\multicolumn{1}{c}{--} \\
& *& [36.1$\pm$9.9] & [24.5$\pm$0.9] & & & & & &\\

& & & & & & & & &\\

Swift J0911.2+4533 & 4.2$\pm$0.2 & $<$0.3& 4.4$\pm$0.2 & Sy2 & 0.0269 & 127.1 & 0.019 & $>$0.360&8.4$\times$10$^{37}$ (1.4 GHz) \\
& [4.4$\pm$0.2] & [$<$ 0.2] & [4.4$\pm$0.3] & & & & & &\\

& & & & & & & & &\\

Swift J0917.2$-$6221  & *& 1450$\pm$40 & 270$\pm$3 & Sy1.2  & 0.057 & 274.3 & 0.182 &--&4.7$\times$10$^{39}$ (843 MHz)  \\
& * & [2380$\pm$60] & [518$\pm$6] & & & & & &\\
& & & & & & & & &\\

& & & & & & & &  &\\

Swift J0923.7+2255 & *& 127$\pm$13 & 92$\pm$4 & NLSy1  & 0.034 & 160.9 & 0.043&--&4.5$\times$10$^{38}$ (1.4 GHz)  \\
& *& [140$\pm$14] & [105$\pm$5] & & & & & &\\

& & & & & & & & &\\

Swift J1049.4+2258 & 26$\pm$1& 4.4$\pm$0.4 & 87$\pm$1 & Sy2  & 0.033 & 156.0 & 0.029 &0.261&\multicolumn{1}{c}{--}\\
& [29$\pm$1]& [4.7$\pm$0.5]& [$93\pm$1] & & & & & &\\
& & & & & & & & &\\

\noalign{\smallskip} 
\hline
\noalign{\smallskip} 
\multicolumn{10}{l}{Note: emission line fluxes are reported both as 
observed and (between square brackets) corrected for the intervening Galactic} \\ 
\multicolumn{10}{l}{absorption $E(B-V)_{\rm Gal}$ along the object line of sight 
(from Schlegel et al. 1998). Line fluxes are in units of 10$^{-15}$ erg cm$^{-2}$ s$^{-1}$,} \\
\multicolumn{10}{l}{The typical error on the redshift measurement is $\pm$0.001 
but for the SDSS and 6dFGS spectra, for which an uncertainty} \\
\multicolumn{10}{l}{of $\pm$0.0003 can be assumed.} \\
\multicolumn{10}{l}{$^*$: heavily blended with [N {\sc ii}] lines} \\
\noalign{\smallskip} 
\hline
\hline
\end{tabular} 
\end{center}
\end{table*}

\section{Results}

In this section we describe the results of the observations reported in the previous sections.
The {\it B} magnitudes and the redshifts of the objects mentioned below, if not otherwise stated, are extracted from the LEDA archive (Prugniel et al.  2005).
Moreover, for each source we list X-ray, radio and Infrared counterparts in the {\it ROSAT} all sky survey bright source Catalog (Voges et al. 1999), {\it RXTE} slew survey (Revnivtsev et al. 2004), {\it XMM-Newton} serendipitous survey (Watson et al. 2009), NVSS radio Catalog (Condon et al. 1998), SUMSS Catalog (Mauch et al. 2003), MGPS-2 catalog (Murphy et al. 2007) and IRAS point source Catalog (IRAS 1988). 

The optical analysis of the 17 extragalactic objects in our sample reveal that 16 are AGNs and 1 is an X-ray bright optically normal galaxy (XBONG), that is a `passive' galaxy with absorption lines in the optical spectrum (Comastri et al. 2002). The results of our optical study are reported in Table \ref{res} where we list for each source the H$_{\alpha}$, H$_{beta}$ and [O{\sc iii}] fluxes, the classification,  the redshift estimated from the narrow lines, the luminosity distance given in Mpc, the Galactic color excess and the color excess local to the AGN host and finally the radio luminosity.

The results of the X-ray analysis are reported in Table \ref{X} where we list the X-ray coordinates, the values of the spectral parameters 
obtained together with the values of the Galactic column density, the hard X-ray (14-195 keV) flux collected from the literature and the luminosity in the energy ranges 2-10 keV and 14-195 keV.
We point out that generally the values of the spectral parameters found with our analysis are in good agreement with those
reported before (e.g. Tueller et al. 2009 and Winter et al. 2008).
\vspace{-0.1mm}
We would like to point out that all previous classifications of our sample sources reported in the literature 
are not supported by any published optical spectrum. 
Only one source, Swift J0505.7$-$2348, has a published spectrum, which is however
normalized to the continuum level optical spectrum (Bikmaev et al. 2006): in this work we have performed a deeper study on the source using a flux-calibrated optical spectrum. 

%The main observed parameters for each source are listed in Table \ref{res}, 
Below we present the main results on the objects of our sample.

%\addtocounter{figure}{-1}
\begin{figure*}[th!]
%\begin{center}
\hspace{-.1cm}
\centering{\mbox{\psfig{file=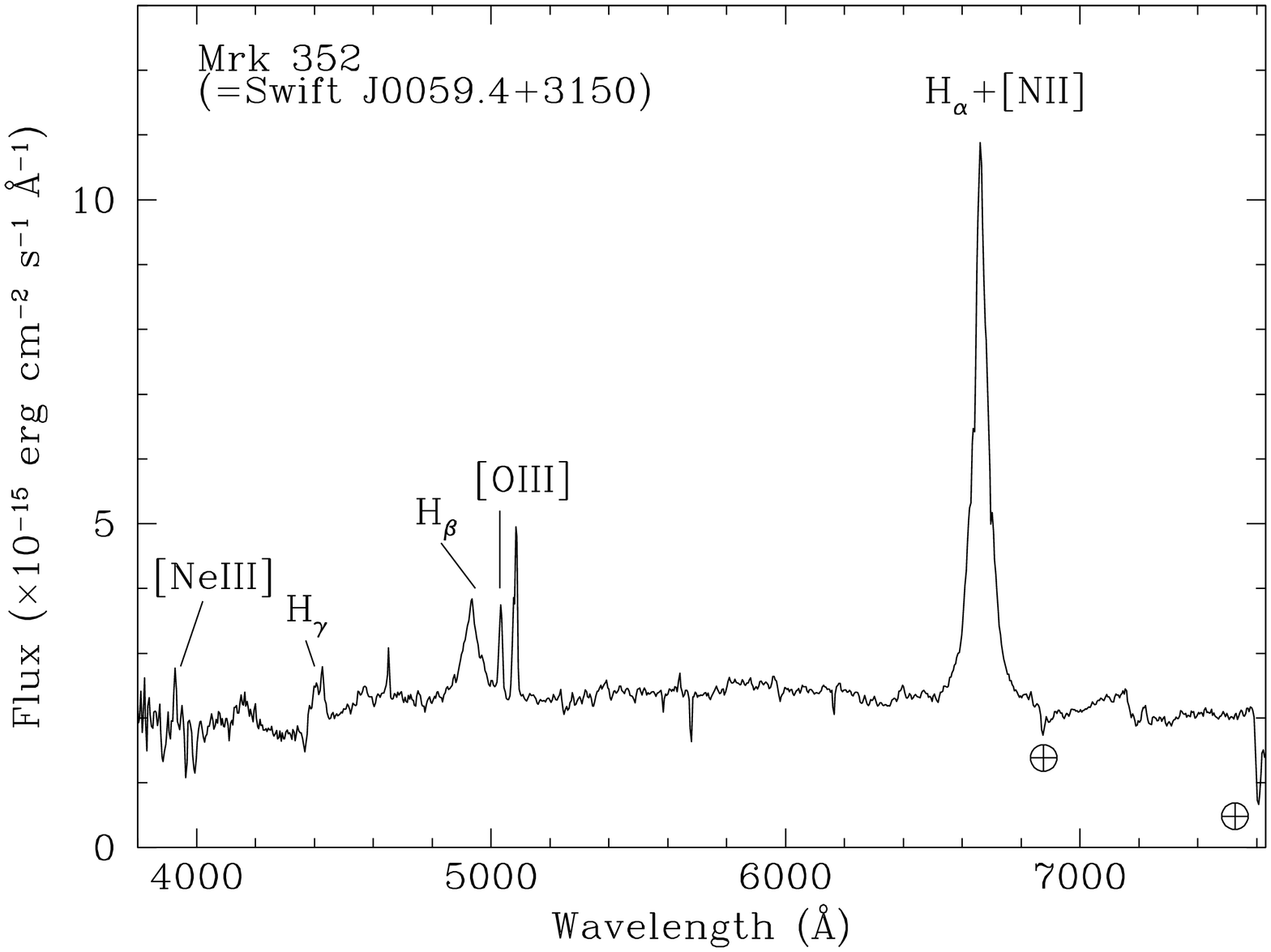,width=9cm,angle=270}}}
%\vspace{-.3cm}
\centering{\mbox{\psfig{file=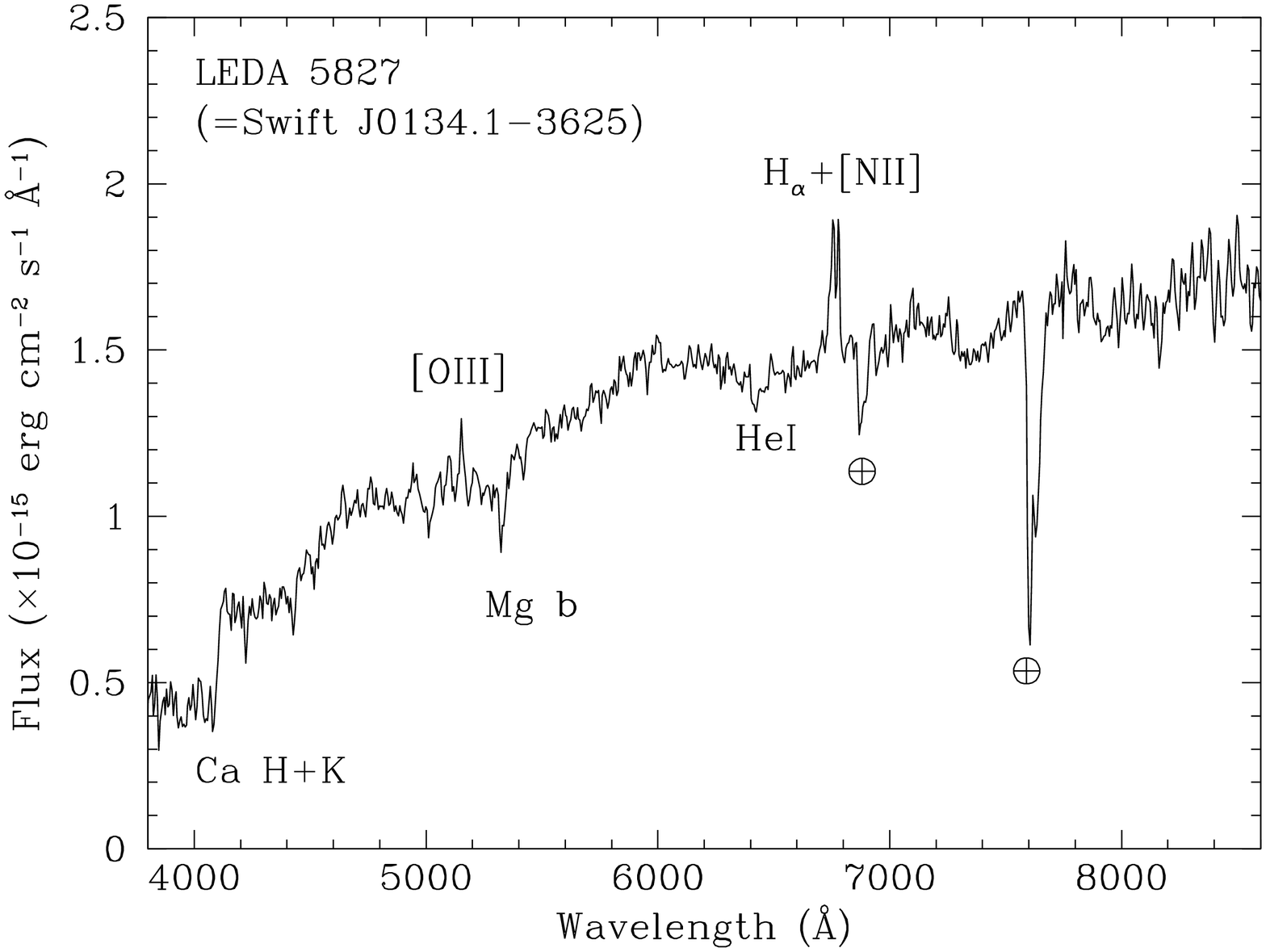,width=9cm,angle=270}}}
\centering{\mbox{\psfig{file=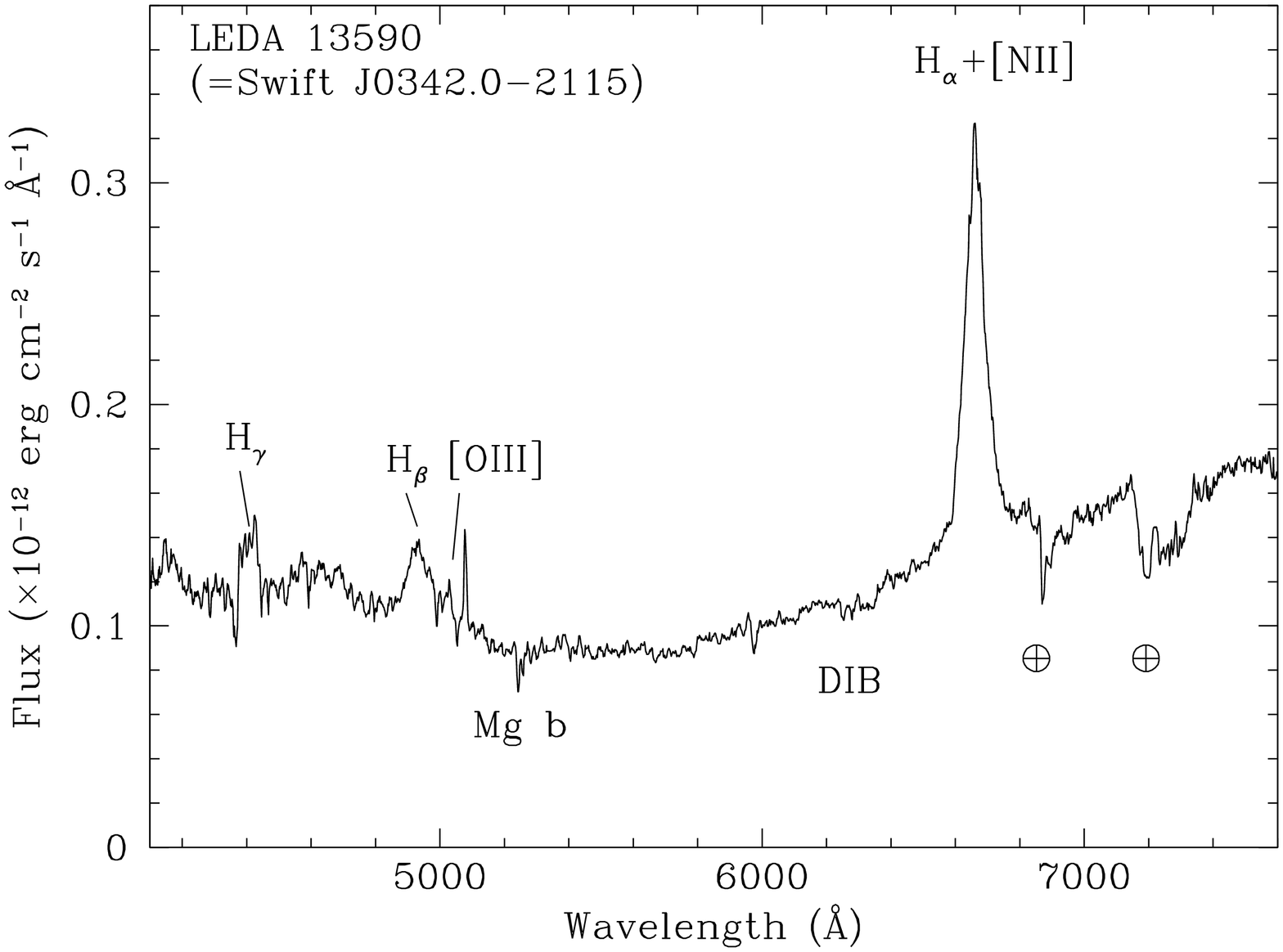,width=9cm,angle=270}}}
\centering{\mbox{\psfig{file=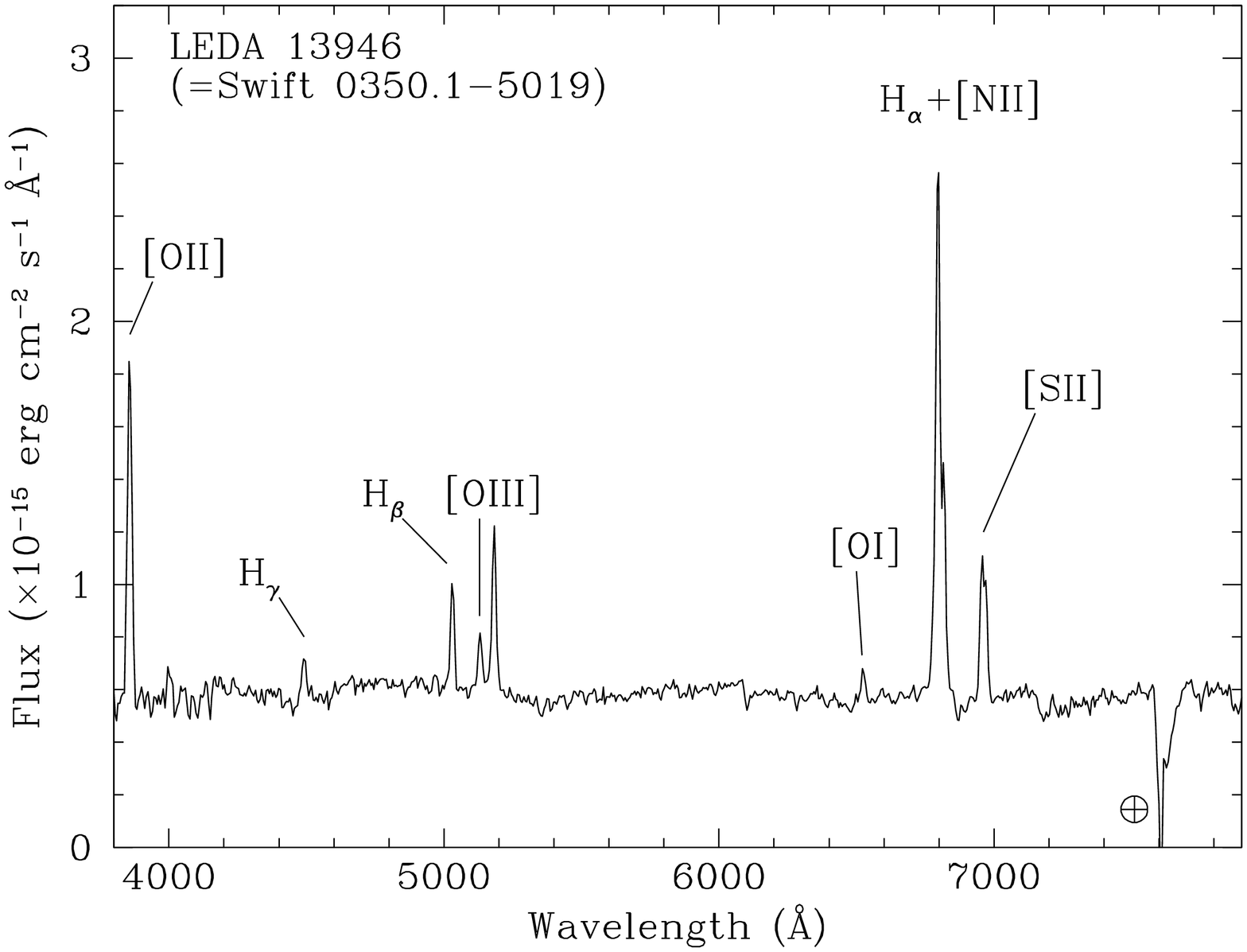,width=9cm,angle=270}}}
\centering{\mbox{\psfig{file=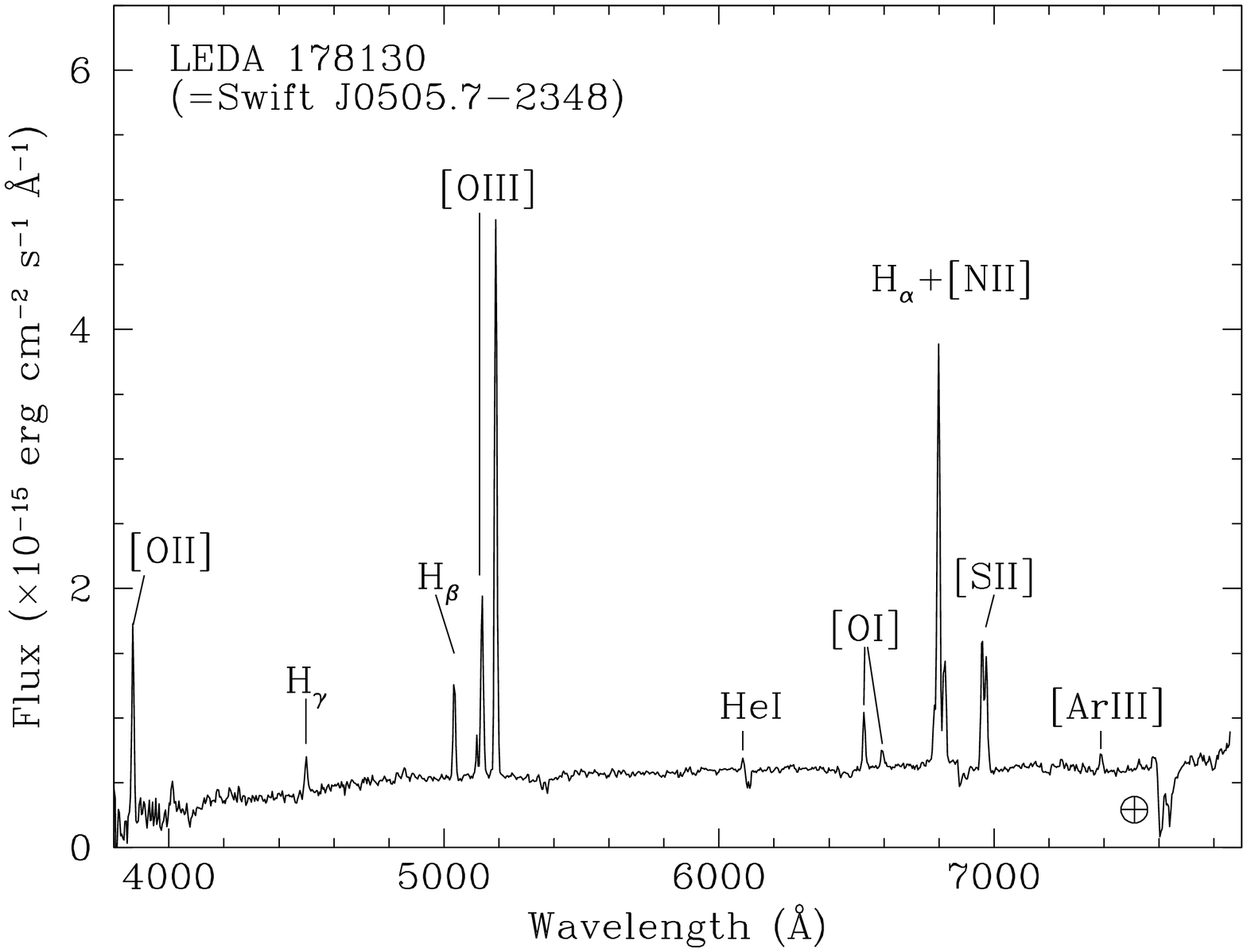,width=9cm,angle=270}}}
%\vspace{-.3cm}
\centering{\mbox{\psfig{file=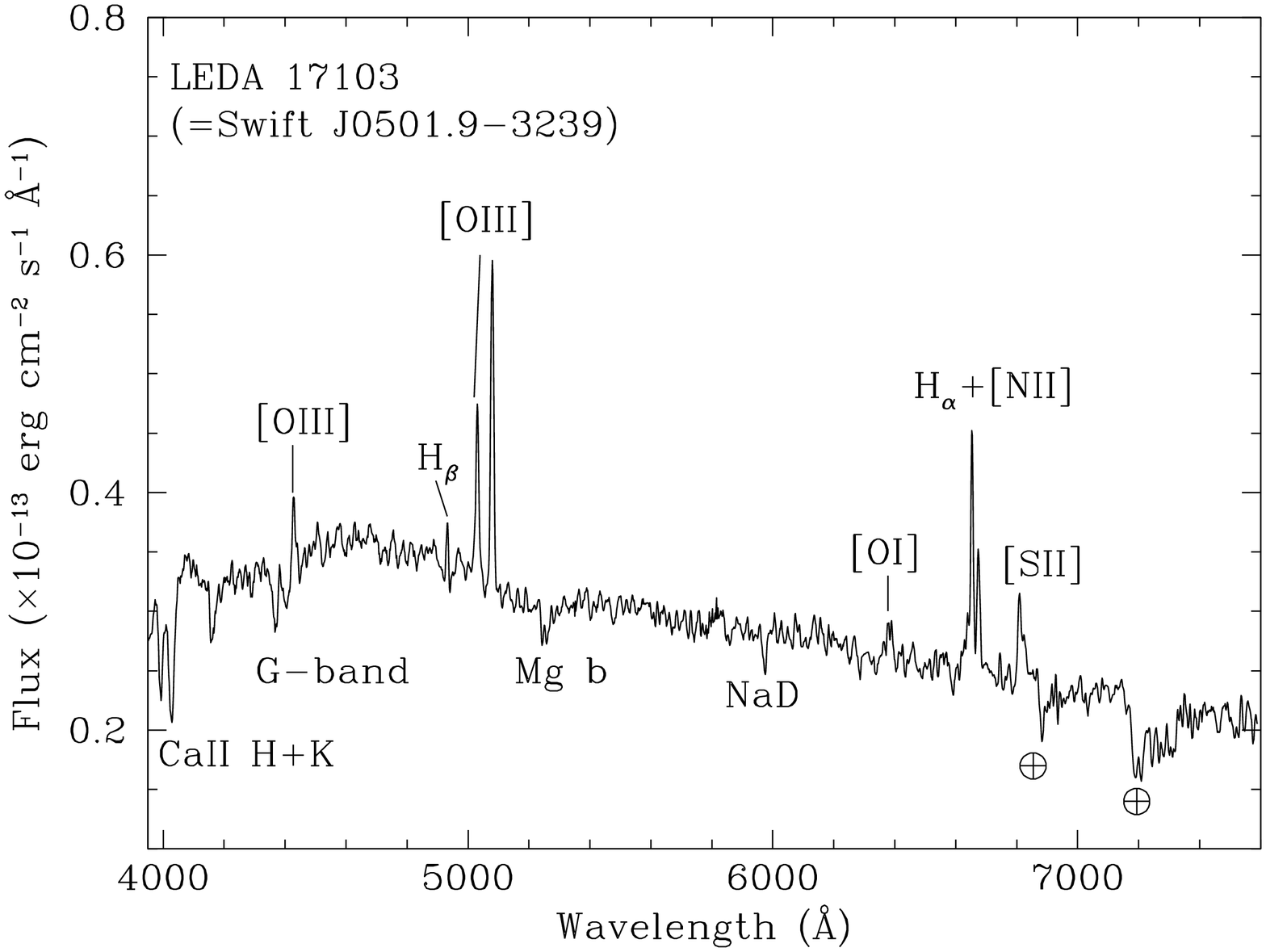,width=9cm,angle=270}}}
%\vspace{-.3cm}

%\vspace{-.3cm}
\caption{Spectra (not corrected for the intervening Galactic absorption) of the 
optical counterpart of  Swift J0059.4$+$3150, Swift J0134.1-3625, Swift J0342.0-2115, Swift J0350.1-5019, Swift J0505.7$-$2348 and
Swift J0501.9$-$3239.
}\label{spectra}
%\end{center}
\end{figure*}
\begin{figure*}[th!]
%\begin{center}
\hspace{-.1cm}
\centering{\mbox{\psfig{file=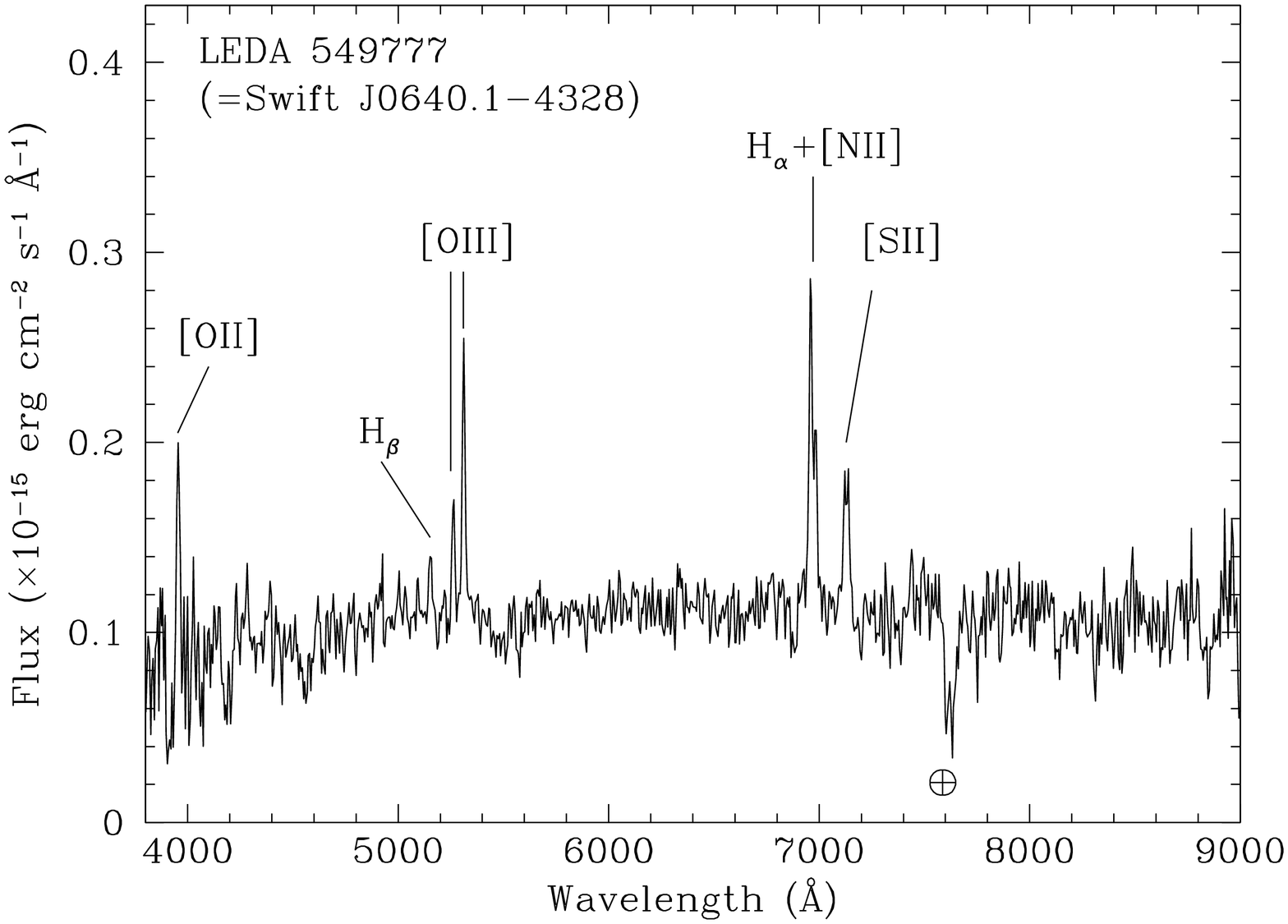,width=9cm,angle=270}}}
%\centering{\mbox{\psfig{file=sj0641.3_3257.ps,width=9cm,angle=270}}}
\centering{\mbox{\psfig{file=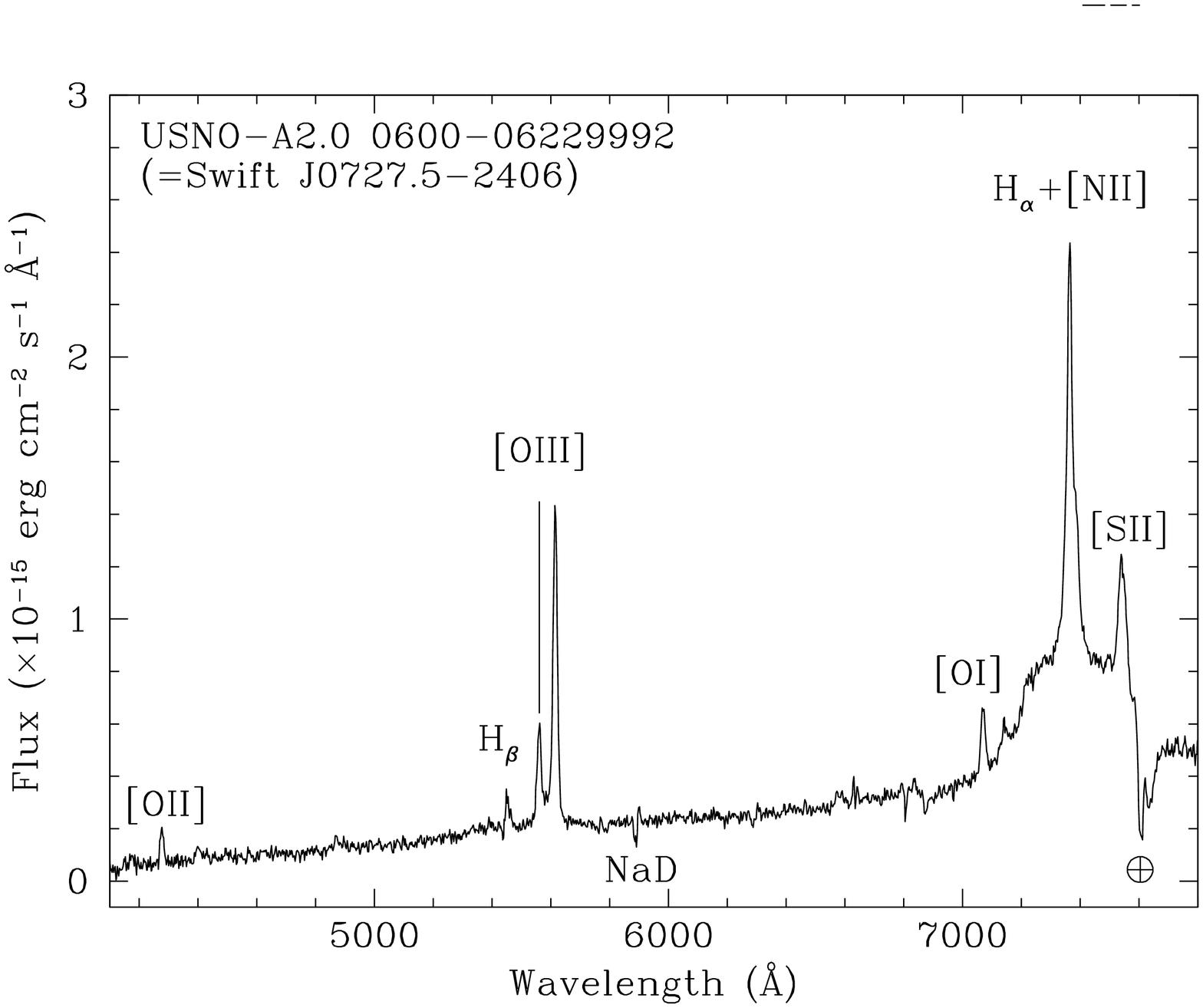,width=9cm,angle=270}}}
\centering{\mbox{\psfig{file=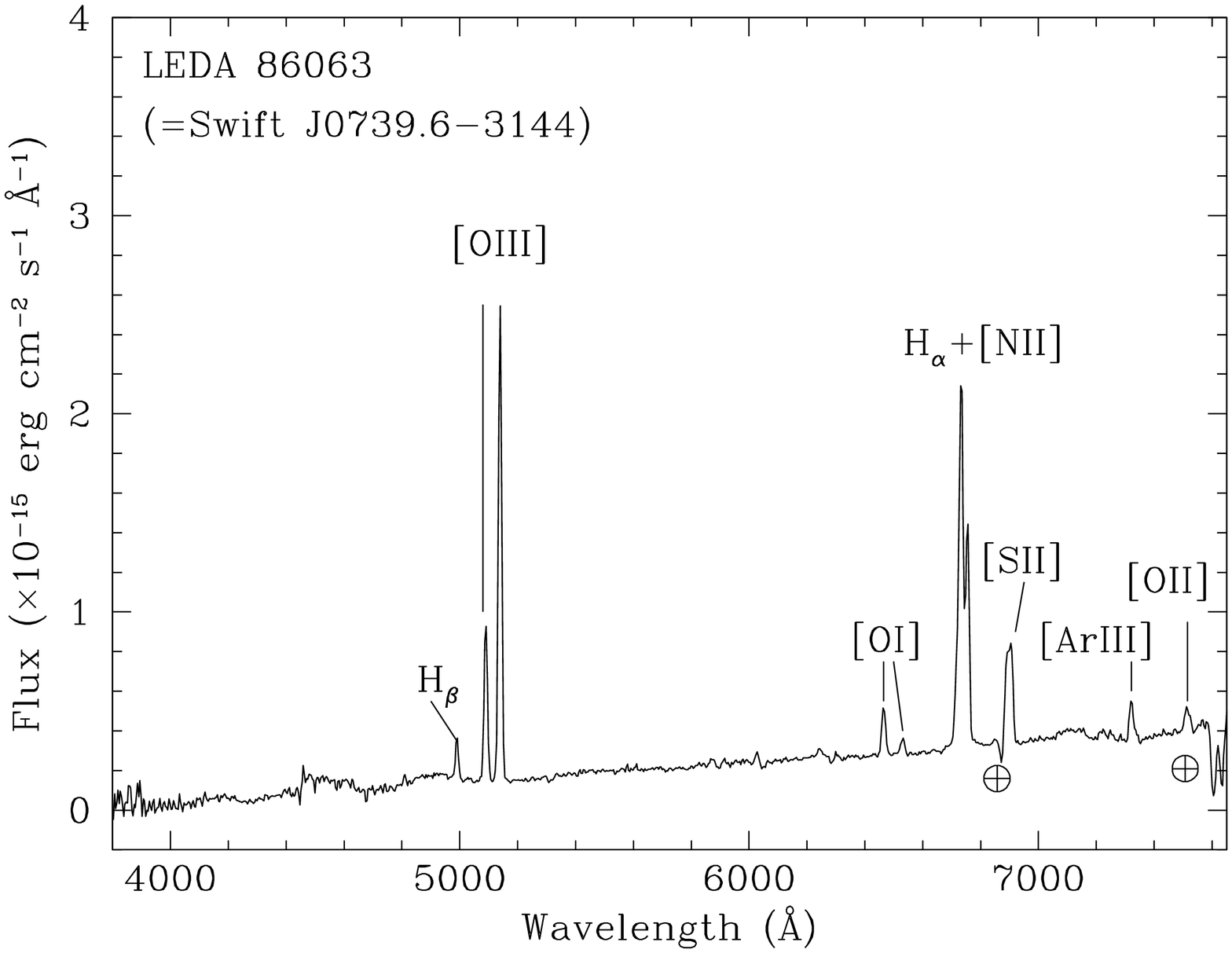,width=9cm,angle=270}}}
\centering{\mbox{\psfig{file=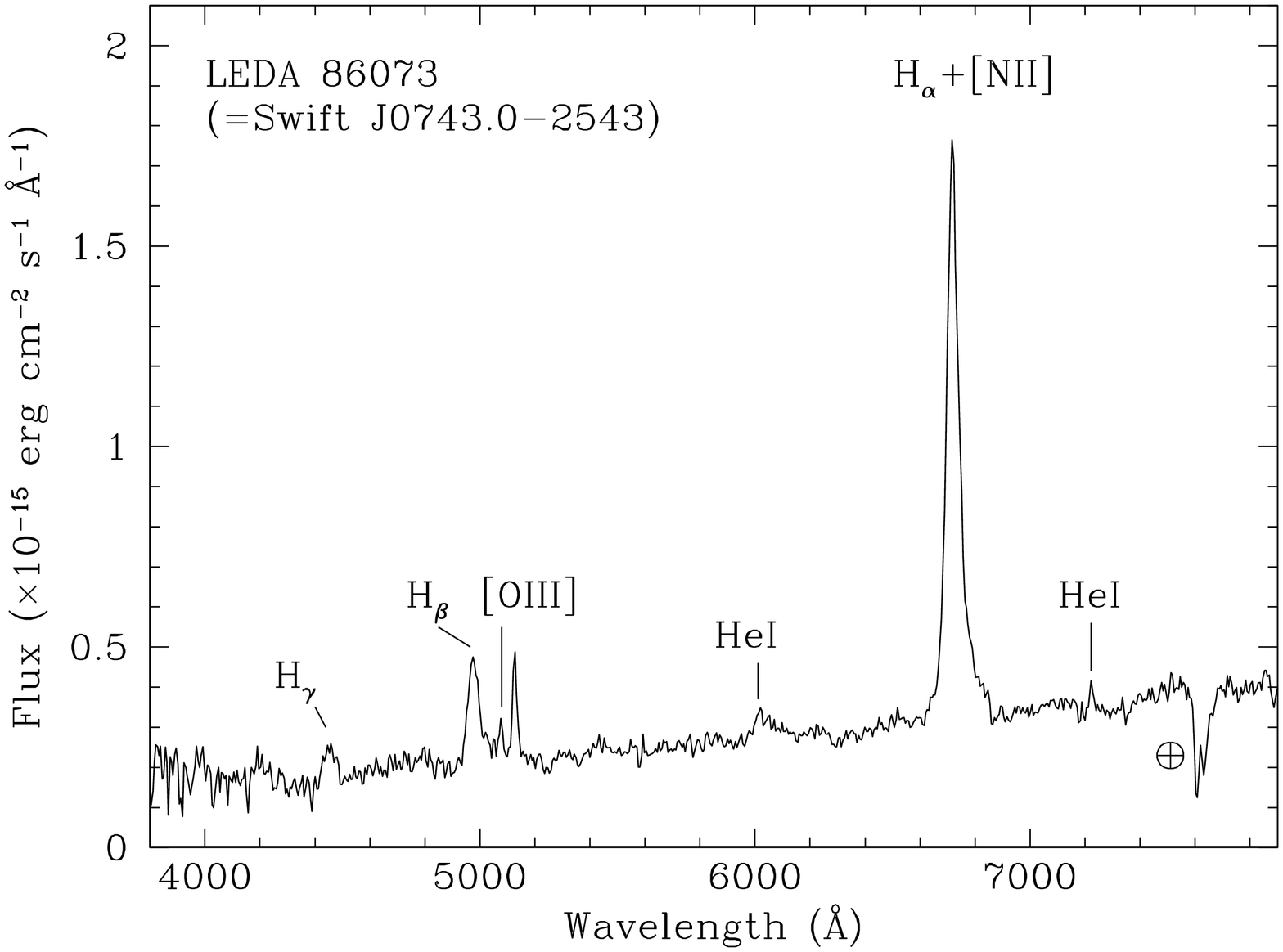,width=9cm,angle=270}}}
%\vspace{-.3cm}
%\vspace{-.3cm}
\centering{\mbox{\psfig{file=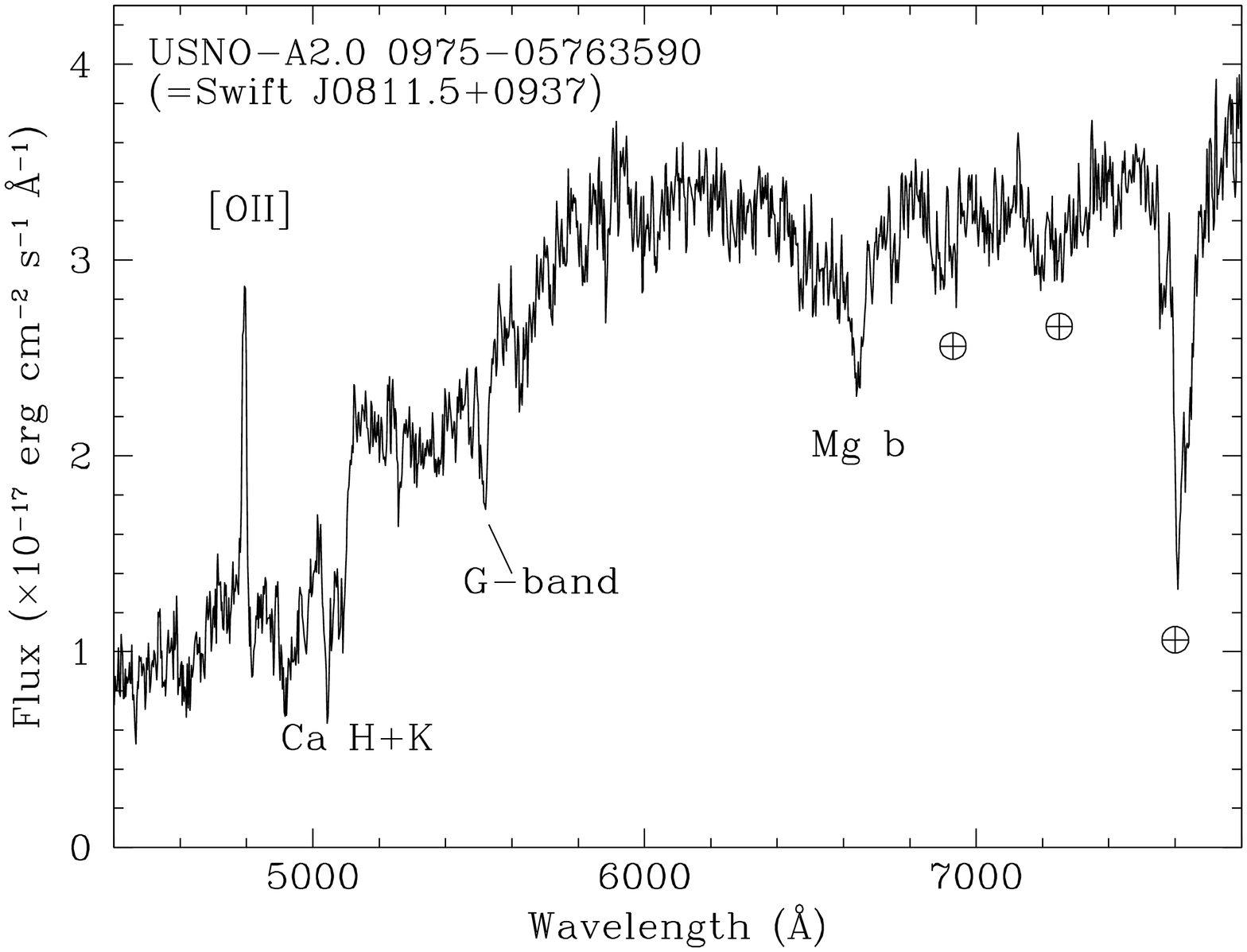,width=9cm,angle=270}}}
\centering{\mbox{\psfig{file=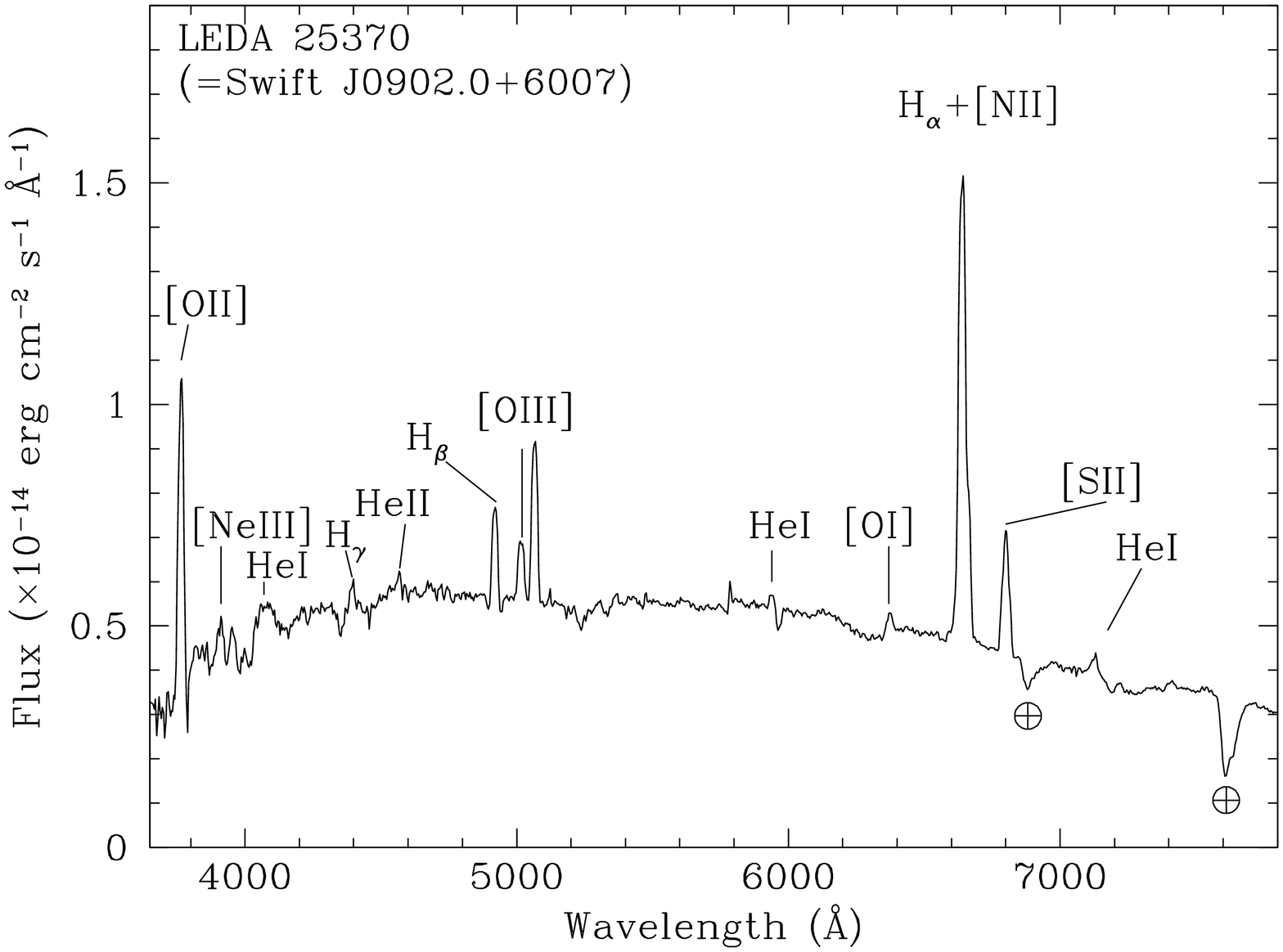,width=9cm,angle=270}}}
%\vspace{-.3cm}

%\vspace{-.3cm}
%\vspace{-.3cm}
\caption{Spectra (not corrected for the intervening Galactic absorption) of the 
optical counterpart of Swift J0640.1$-$4328, Swift J0727.5$-$2406, Swift J0739.6$-$3144, Swift J0743.0$-$2543 and Swift J0811.5+0937, Swift J0902.0+6007. 
}\label{spectra2}
%\end{center}
\end{figure*}

\begin{figure*}[th!]
%\begin{center}
\hspace{-.1cm}
%\centering{\mbox{\psfig{file=sj0854.7_2.ps,width=9cm,angle=270}}}
\centering{\mbox{\psfig{file=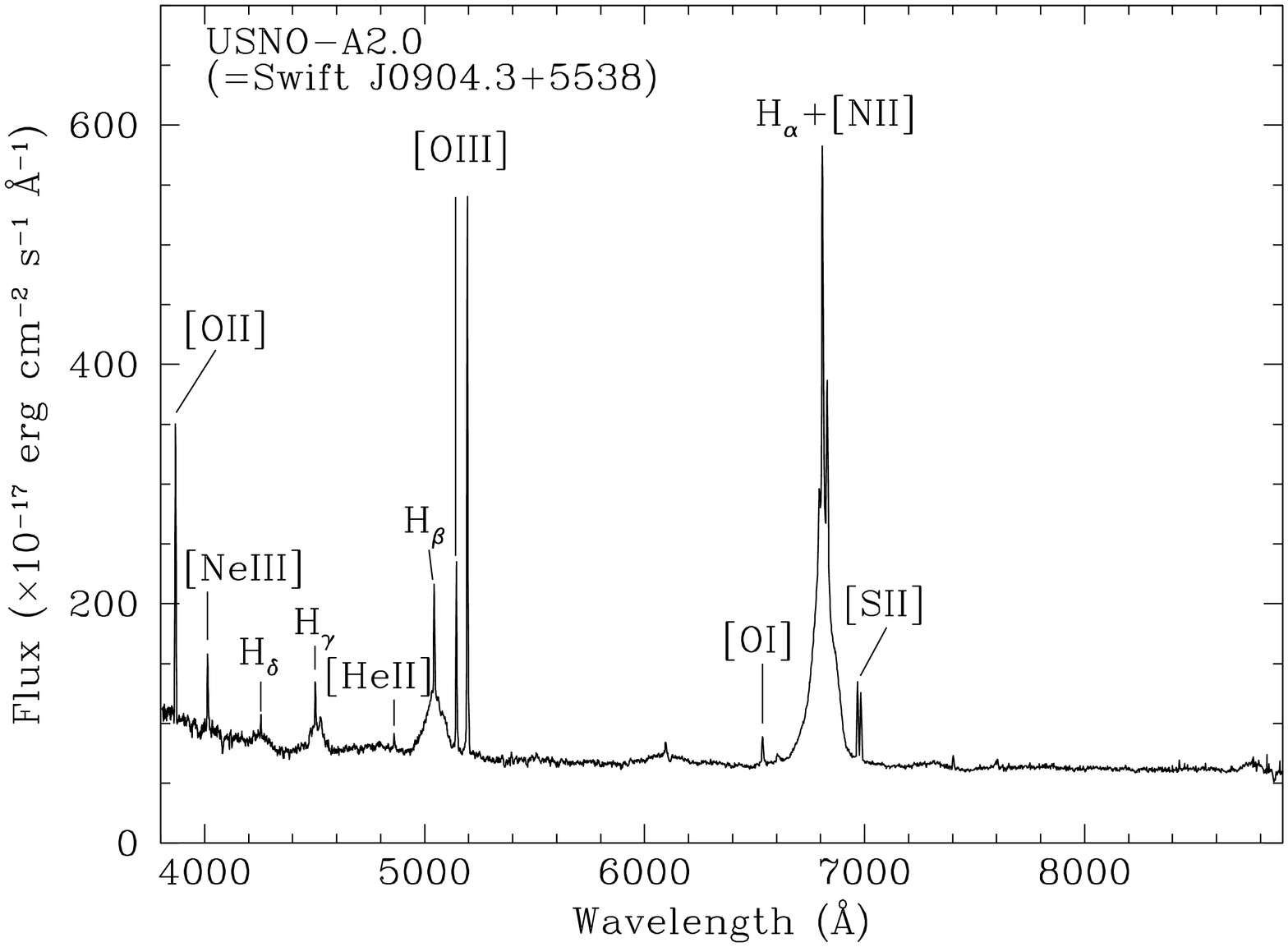,width=9cm,angle=270}}}
\centering{\mbox{\psfig{file=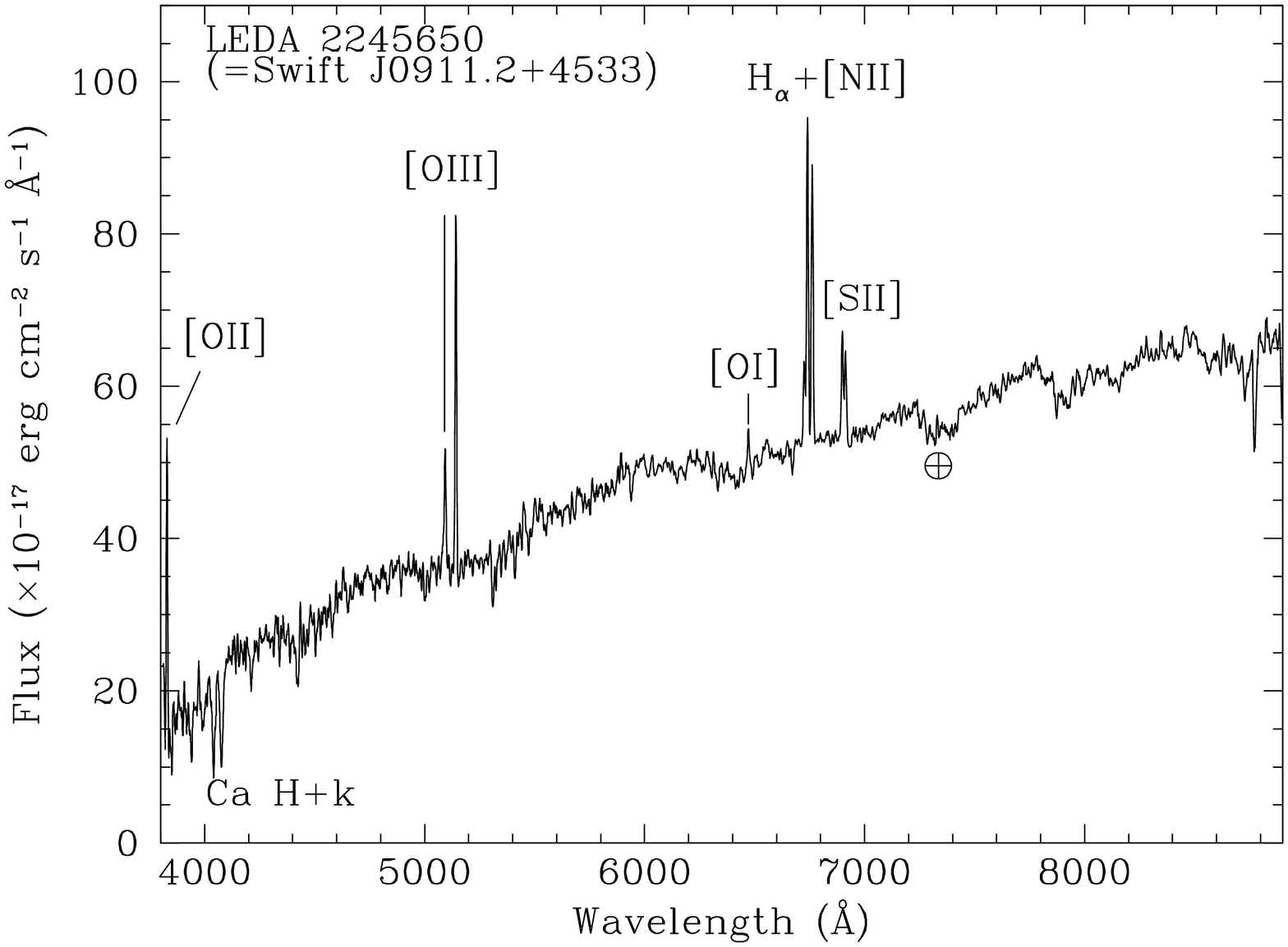,width=9cm,angle=270}}}
\centering{\mbox{\psfig{file=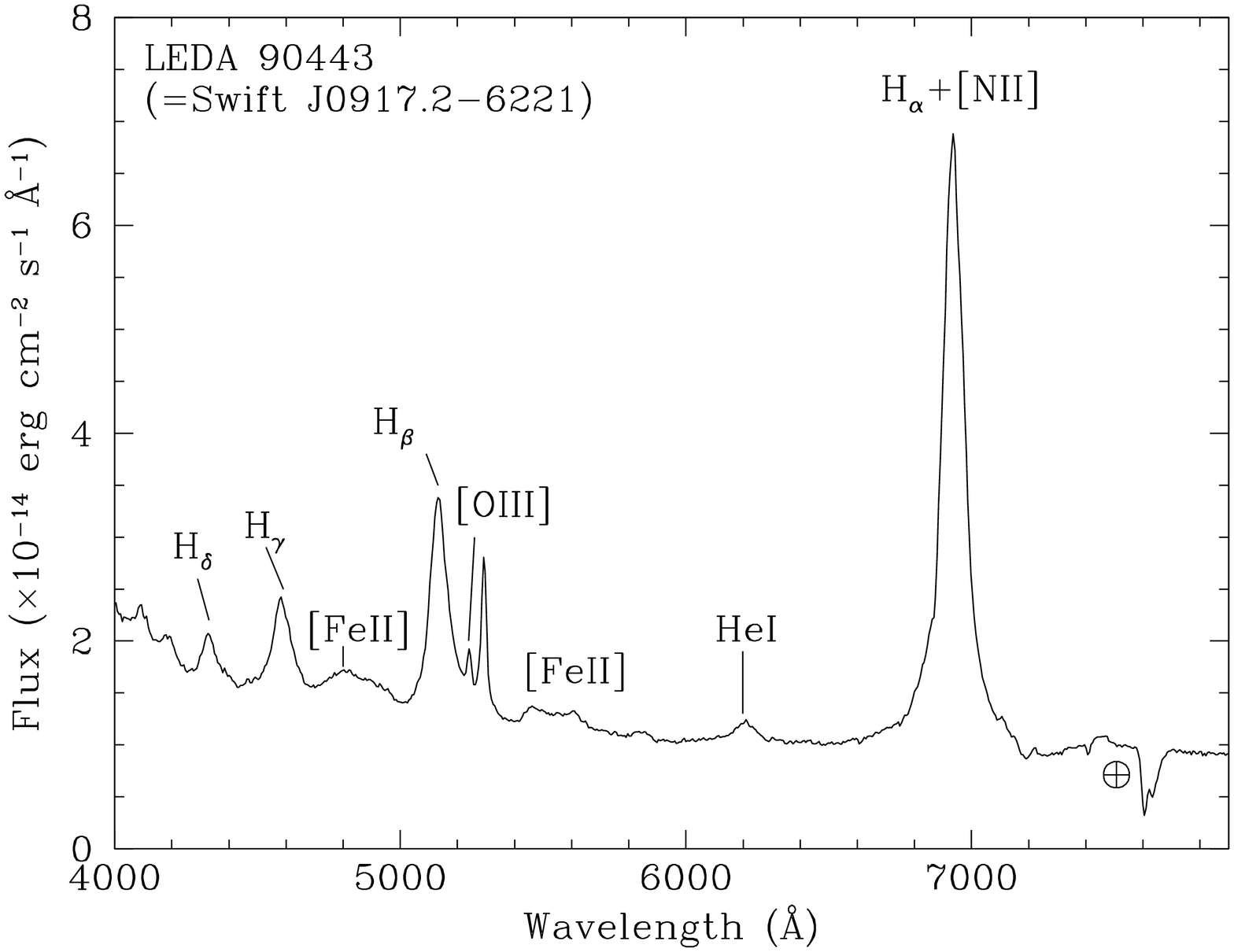,width=9cm,angle=270}}}
\centering{\mbox{\psfig{file=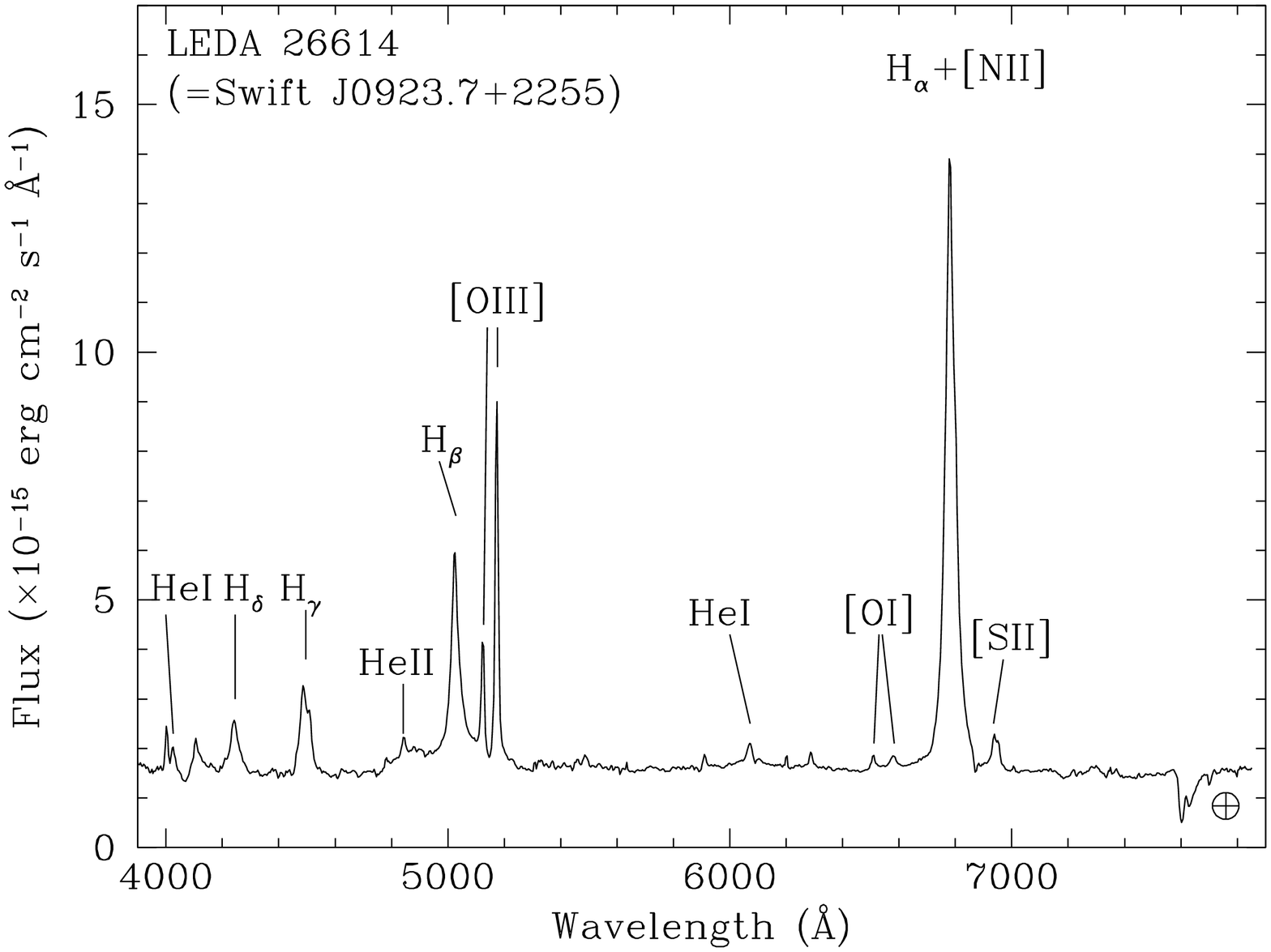,width=9cm,angle=270}}}
\parbox{9cm}{\psfig{file=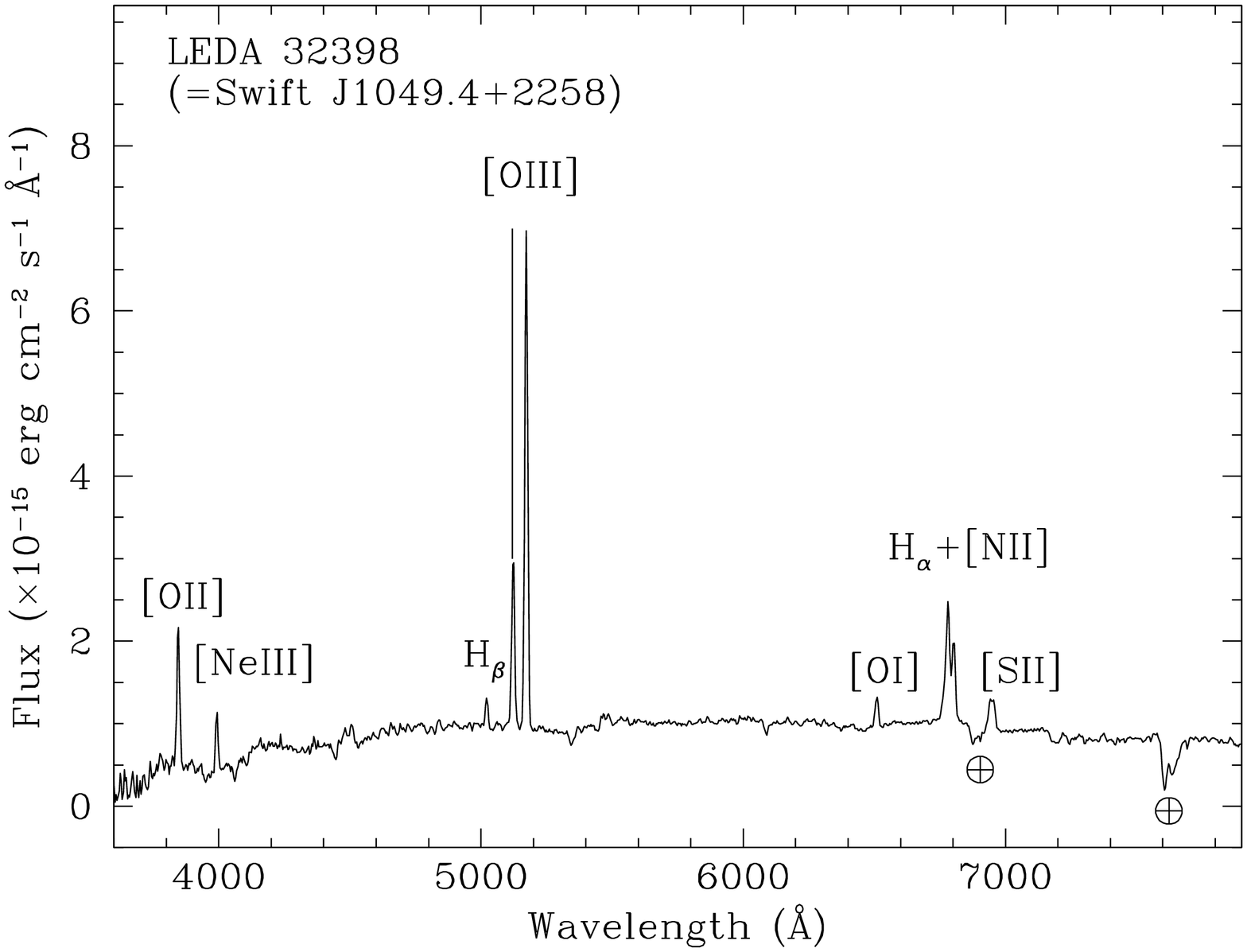,width=9cm,angle=270}}
%\vspace{-.3cm}
%\vspace{-.3cm}
%&\centering{\mbox{\psfig{file=specsumswij0811_trim.ps,width=9cm,angle=270}}}
%\vspace{-.3cm}
%\centering{\mbox{\psfig{file=sj0854.7_2.ps,width=9cm,angle=270}}}
%\vspace{-.3cm}
%\centering{\mbox{\psfig{file=specsumj0902_6007.ps,width=9cm,angle=270}}}
%\vspace{-.3cm}
%\centering{\mbox{\psfig{file=swij0917.2_6221.ps,width=9cm,angle=270}}}
%\centering{\mbox{\psfig{file=specsummcg0422042.ps,width=9cm,angle=270}}}
\hspace{1.3cm}
\parbox{7.5cm}{
\vspace{-0.5cm}
\vspace{-0.5cm}
\caption{Spectra (not corrected for the intervening Galactic absorption) of the 
optical counterpart of Swift J0904.3+5538, Swift J0911.2+4533, Swift J0917.2-6221 and Swift J0923.7+2255, Swift J1049.4+2258.  
}\label{spe}}
%\end{center}
\end{figure*}

\begin{table*}[tb]
%\begin{center}
\caption[]{Main results obtained from the analysis of the X-ray spectra of the 17 sources of the present 
sample.}\label{X}
\scriptsize
\resizebox{18cm}{!}{
\begin{tabular}{lcccccccccr}
\noalign{\smallskip}
\hline
\hline
\noalign{\smallskip}
\multicolumn{1}{c}{Source} & \multicolumn{1}{c}{RA} &\multicolumn{1}{c}{Dec} &N$_H$ Gal & N$^{\bullet}_H$  &  $\Gamma^{+}$  & F$_{(2-10)keV}$ &F$_{(14-195)keV}$& $L_{\rm X}$&Ref.$^{*}$ \\
%\cline{3-6}
\noalign{\smallskip}
& & & $\times 10^{22}$cm$^{-2}$   &$\times 10^{22}$cm$^{-2}$  & &$\times 10^{-12}$erg s$^{-1}$ cm$^{-2}$&$\times 10^{-11}$erg s$^{-1}$ cm$^{-2}$& 10$^{43}$ erg s$^{-1}$ &\\
\noalign{\smallskip}
\hline
\noalign{\smallskip}

Swift J0059.4+3150$^{A}$ &00 59 53.29  & +31 49 36.6 &0.055& --& $1.67^{+0.03}_{-0.03}$ & 7.3 & 4.2& 0.6 (2--10)&1\\ 
& & & & & & & & 3.1 (14--195)&\\
 
& & & & & & & & &\\

Swift J0134.1$-$3625$^{A}$ & 01 33 57.71 & $-$36 29 38.4 &0.019& $75^{+34}_{-21}$ & [1.8]  & 1.9& 5.4& 0.8 (2--10)&1  \\
& & & & & & & &7.2 (14-195) & \\

& & & & & & & & & \\

Swift J0342.0$-$2115$^{A}$  &03 42 03.62 & $-$21 14 39.6 &0.023 & --& $1.94^{+0.05}_{-0.05}$&23.8&4.6&$\sim$1.3 (2--10)&1 \\
& & & & & & & &1.8 (14--195) &\\
& & & & & & & & &\\

Swift J0350.1$-$5019$^{A}$  &03 50 23.08 &$-$50 18 11.4  &0.012& $17^{+9}_{-7}$ & [1.8]& 3.1&3.1&1.0 (2--10) &1\\ 
& & & & & & & &9.5 (14--195) &\\

Swift J0505.7$-$2348$^{A}$ &05 05 45.70 &$-$23 51 13.3 & 0.021&$6^{+0.5}_{-0.5}$& $1.4^{-0.4}_{+0.3}$& 15.5 &7.2&5.3 (2--10)&1 \\

& & & & & & & &21.6 (14--195) &\\
& & & & & & & & &\\

Swift J0501.9$-$3239$^{A}$&05 19 35.82  &$-$32 39 26.6  & 0.018&--& 1.45$^{+0.05}_{-0.05}$ &13.6&6.2&0.5 (2--10)&1 \\

& & & & & & & &1.2 (14--195) &\\
& & & & & & & & &\\
Swift J0640.1$-$4328$^{A}$&06 40 37.89  & $-$43 21 19.8& 0.025 &11.7$^{+4.5}_{-3.6}$ &[1.8]& 2.2&2.0&2.3 (2--10)&1\\

& & & & & & & & 28.9 (14--195)&\\
& & & & & & & & &\\
% Swift J0641.3+3257& in abs.& in abs. & $<$3.7 & XBONG & 0.016 & 74.7 & 0.149 &-- & 22.53 (2--10) \\
% & [in abs.]& [in abs.] & [$<$8.5] & & & & & & 3.67 (14--195) \\
%& & & & & & & & & \\
Swift J0727.5$-$2406$^{C}$ & 07 27 21.05& $-$24 06 32.2 &0.650&--& --&0.6&2.2& 2.7 (2--10)&2 \\
& & & & & & & & 93.2 (14--195)&\\
& & & & & & & & &\\
& & & & 47$^{+13}_{-12}, { [0.7]}$ & & & & &\\
Swift J0739.6$-$3144$^{B}$ &07 39 44.61& $-$31 43 01.6& 0.431 &  & { 1.77$^{+0.07}_{-0.09}$} &0.9&2.5&0.2 (2--10)&2\\
& & & & 14$^{+2.3}_{-1.9},{ [0.9]}$ & & & &4.5 (14--195) &\\
& & & & & & & & &\\
& & & & & & & & &\\

Swift J0743.0$-$2543$^{B}$&07 43 14.95 &$-$25 45 44.3  & 0.581& -- & --& 14.7$^{\triangle}$  &2.0& 2.1 (2--10)&2\\

& & & & & & & &2.8 (14--195) &\\

& & & & & & & & &\\

Swift J0811.5+0937$^{C}$&08 11 30.99& +09 33 51.6 &0.024 &--& -- & 0.1&2.6&3.0 (2--10) &2 \\
& & & & & & & &778.1 (14--195) &\\

& & & & & & & & &\\
%Swift J0854.7+1502  & 7.1$\pm$1.2& 0.9$\pm$0.1 & 4.9$\pm$0.3 & Sy2  & 0.071 & 345.0 & 0.040 &0.306 & 24.4 (14--195) \\
% & [7.2$\pm$1.2]& [1.1$\pm$0.2] & [5.5$\pm$0.3] & & & && &  \\
% & & & & & & & & & \\
Swift J0902.0+6007$^{A}$& 09 01 58.81 &+60 09 05.8  & 0.043 & 8$^{+8.6}_{-4.0}$& [1.8]  & 1.1 &3.1&0.04 (2--10)&1\\

& & & & & & & & 1.2 (14--195)&\\

& & & & & & & & &\\

Swift J0904.3+5538$^{A}$& 09 04 36.92  & +55 35 59.7&0.022& --& 1.44$^{+0.07}_{-0.07}$&4.9&  1.9&1.8 (2--10)&1\\

& & & & & & & & 7.2 (14--195)&\\

& & & & & & & & &\\

Swift J0911.2+4533$^{A}$&09 11 29.98  &+45 28 04.1 & 0.012& 30$^{+9.5}_{-7}$&[1.8] &2.0&1.8&0.4 (2--10)&1  \\

& & & & & & & &5.8 (14--195) &\\

& & & & & & & & &\\

Swift J0917.2$-$6221$^{A}$&09 16 09.14&$-$62 19 28.4  &0.158 & 0.7$^{+0.3}_{-0.3}$ & 1.67$^{+0.2}_{-0.2}$  &16.0&3.3&14.8 (2--10)&1  \\

& & & & & & & &28.8 (14--195) &\\

& & & & & & & & &\\

Swift J0923.7+2255$^{A}$&09 23 42.89&+22 54 33.2 & 0.031& -- & 1.95$^{+0.05}_{-0.05}$  &11.8& 4.5&3.7 (2--10)&1 \\

& & & & & & & &12.7 (14--195) &\\

& & & & & & & & &\\

Swift J1049.4+2258$^{A}$ &10 49 30.88& +22 57 52.9 & 0.019 & 20$^{+5}_{-3}$ & [1.8]& 0.7&3.7 &0.2 (2--10)&1 \\
& & & & & & & &10.9 (14--195) &\\

& & & & & & & & &\\

\noalign{\smallskip} 
\hline
\noalign{\smallskip} 
\hline
\hline
\multicolumn{10}{l}{$^+$ The square brackets in the $\Gamma$ column indicate that we used a fixed value}&\\
\multicolumn{10}{l}{$^*$ This column reports the references for the hard X-ray flux}&\\
\multicolumn{10}{l}{$^{\triangle}$ Source fluxes were calculated from count rates (see {\tt http://heasarc.gsfc.nasa.gov/db-perl/W3Browse/w3table.pl})}&\\
\multicolumn{10}{l}{$^{\bullet}$ The values in the square brackets indicate the covering fraction}&\\
\multicolumn{10}{l}{$^1$ Tueller et al. (2009) }&\\
\multicolumn{10}{l}{$^2$ Ajello et al. (2007) }&\\
\multicolumn{10}{l}{$^{A}${ Soft X-ray data from} {\it Swift}/XRT}&  \\
\multicolumn{10}{l}{$^{B}${ Soft X-ray data from} {\it XMM}/EPIC}& \\
\multicolumn{10}{l}{$^{C}${ Soft X-ray data from} {\it Chandra}}& \\
\hline
\hline
\end{tabular}}
%\end{center}
\end{table*}

\subsection {Swift J0059.4+3150}
This source, identified as Mrk 352, is associated with the galaxy LEDA 3575 with magnitude B=14.93 and redshift of 0.0149.
It is associated also with the {\it ROSAT} bright source 
1RXS J005953.3+314934, the {\it XMM-Newton} serendipitous
source 2XMM J005953.2+314937 and with the IRAS point source IRAS 00572+3134.

Here, we publish for the first time an optical spectrum of Swift J0059.4+3150 (Fig. \ref{spectra}, upper left panel). It shows a weak continuum, with broad H$_\alpha$+[N{\sc ii}] complex in emission, H$_{\beta}$ and H$_{\gamma}$ emission lines, and the [O{\sc iii}]$\lambda$5007 forbidden narrow emission line. These allow us to derive the redshift of 0.015$\pm$0.001, consistent with the redshift determination reported in the Hyperleda archive.
Thus, we refine the Seyfert 1 AGN classification reported by Tueller et al. (2008) to Seyfert 1.2. 
\subsection{Swift 0134.1-3625}
This object is associated with the galaxy LEDA 5827 with a magnitude B=14.04 and a redshift 0.0299.
Moreover, it is listed in the 2nd {\it XMM} serendipitous source catalog as 2XMMi J013357.6-362935 and in the IRAS Point Source Catalog as IRAS 03398-2124.

The optical spectrum of this source (Fig. \ref{spectra}, upper right panel) shows a red stellar continuum, with a weak narrow emission H$_{\alpha}$+[N{\sc ii}] complex and a weaker forbidden [O{\sc iii}] emission line. The lines have a redshift of 0.029$\pm$0.001, consistent with that reported in the Hyperleda archive.
We classify for the first time this source as a Seyfert 2 galaxy. 
Our classification is more thorough with respect to that given by SIMBAD (in which the source is reported as `radio galaxy') and in Winter et al. (2008) who reported it as `galaxy'.

%Because of the lack of the H$_{\beta}$ emission line we cannot use the ratio H$_{\alpha}$/H$_{\beta}$ to estimate the intrinsic absorption of the AGN, therefore we cannot correct the [O{\sc iii}] flux and so we are not able to estimate the Compton nature through the 2--10 keV/[O{\sc iii}] flux ratio. However, using the F$_{(2-10)keV}$/F$_{(20-100)keV}$ flux ratio ($\sim 0.08$) we assess that this source is Compton thin.
\subsection{Swift J0342.0-2115}
This {\it Swift} source is associated with the galaxy LEDA 13590 with magnitude B=13.31 and a redshift of 0.0145. It is associated with a {\it ROSAT} bright source, an {\it XMM} serendipitous source (1RXS  J034203.8-211428 and 2XMM J034203.6-211439 respectively) and with an IRAS point source, IRAS 03398$-$2124.

Its optical spectrum (Fig. \ref{spectra}, central left panel) shows the emission H$_{\alpha}$+[N{\sc ii}] complex, with the presence of the broad emission H$_{\beta}$ as well as 
the narrow [O{\sc iii}] forbidden emission lines. These lines allow us to derive a redshift of 0.0139$\pm$0.0003, somewhat in agreement with that reported in the Hyperleda archive.
Here, we classify for the first time this source as a Seyfert 1 AGN. 
\subsection{ Swift J0350.1-5019} 
The optical counterpart of this {\it Swift} source is the galaxy LEDA 13946 with magnitude B=16.19.\\
Its optical spectrum (Fig. \ref{spectra}, central right panel) shows a weak continuum with obvious emission lines; a narrow H$_\alpha$+[N{\sc ii}] complex,
H$_{\beta}$ and H$_{\gamma}$ permitted narrow lines, [O{\sc i}]$\lambda$6300, [O{\sc iii}]$\lambda$5007 and
[O{\sc ii}]$\lambda$3727 forbidden narrow lines.
From the features above we are able to classify this source for the first time as a Seyfert 2 AGN at redshift z = 0.035$\pm$0.001.
Our classification is more thorough with respect to that of Tueller et al. (2008) who report it as a `galaxy'. 
%From the measured 2--10 keV/[O{\sc iii}] flux ratio of $\sim$ 130, we can state that this is a Compton thin source. 
%Likewise, if we use the method of Malizia et al. (2007), we again find that this source is Compton thin.
\subsection{ Swift J0505.7-2348}
This source is associated with the galaxy LEDA 178130, with magnitude B=16.76 and redshift of 0.035. 
It is also associated with a NVSS radio source (NVSS 050545$-$235114) with a flux of 7.9$\pm$0.5 mJy at 1.4 GHz. 

Its spectrum (Fig. \ref{spectra}, bottom left panel) shows a very narrow emission H$_\alpha$+[N{\sc ii}] complex with H$_{\beta}$ and H$_{\gamma}$ permitted narrow emission lines, [O{\sc i}]$\lambda$6300, [O{\sc iii}]$\lambda$5007 and [O{\sc ii}]$\lambda$3727 forbidden narrow emission lines.
Therefore, we classify this {\it Swift} source as a Seyfert 2 AGN with z = 0.036$\pm$0.001.
Our results are in agreement with those given by Bikmaev et al. (2006).
\subsection{ Swift J0501.9-3239}
This object is associated with the galaxy LEDA 17103, with magnitude B=13.87 and redshift 0.012337.
Listed in the {\it ROSAT} Bright Source Catalogue as 1RXS J051936.1-323910, this object has a radio counterpart in the NVSS (NVSS 051935-323929) and SUMSS catalogues, with density fluxes at 1.4 GHz and 843 MHz of 14.2 $\pm$1.0 mJy and 20.1$\pm$ 2.0 mJy, respectively. It also shows an infrared counterpart in the IRAS Point Source Catalog named as IRAS 05177-3242.	

As can be see in Fig. \ref{spectra} (lower right panel), its optical spectrum shows narrow H$_{\alpha}$ and H$_{\beta}$ lines in emission, as well as [O {\sc iii}] forbidden emission line. CaII H+K and G-band are also present in absorption.
From the features listed above, we find that Swift J0501.9-3239 is a Seyfert 2 AGN with z = 0.0126$\pm$0.0003, in agreement with the redshift reported in the Hyperleda archive. 
Our classification is thus at variance with that reported in SIMBAD (Seyfert 1 AGN), in Winter et al. (2008) and , who classify it as Seyfert 1.5 AGN. The {\it Swift}/XRT analysis revealed that this source is 
a naked Seyfert 2 AGN (see Table \ref{X}), { that is, a Seyfert 2 with no intrinsic absorption (e.g. Bianchi et al. 2008)} in agreement with the {optical determination of} E(B-V)$_{AGN}$ (see Table \ref{res}) which is substantially lower than the other type 2 AGNs { of our sample}. 

%The thickness parameter of this source is T $\sim$ 30 and the softness parameter of Malizia et al. (2007) is $\sim$0.9; we are thus able to assess the Compton thin nature of this source.
%The radio luminosities are 8.3$\times 10^{37}$ erg s$^{-1}$ at 1.4 GHz and 7.0$\times 10^{37}$ erg s$^{-1}$ at 843 MHz for the measured redshift.  
\subsection{ Swift J0640.1-4328}
The optical counterpart of this source has been identified as LEDA 549777, with a B magnitude of 16.79.
This object is also a radio emitter, being reported in the SUMSS catalog with a flux of 93.4$\pm$3.0 mJy at 843 MHz.

The optical spectrum (Fig. \ref{spectra2}, upper left panel) shows a weak continuum with a low S/N, H$_\alpha$+[N{\sc ii}] emission complex, a weak narrow H$_{\beta}$ and [O{\sc iii}]$\lambda$5007 forbidden emission line.
Using the information extracted from our optical spectrum we are able to classify this object as a Seyfert 2 AGN with a redshift of 0.061$\pm$0.001.
In the Tueller et al. (2008) catalog this object is classified as `galaxy', with no reported redshift. 

\subsection{ Swift J0727.5-2406}
The optical counterpart of this {\it Swift} source is USNO$-$A2.0 0600-06229992 with magnitude R=17.3.
This object, as reported in Ajello et al. (2008), is associated with the {\it ROSAT} faint source 
1RXS J072720.8$-$240629 and with the radio object NVSS J072721$-$240632 with a flux density of \mbox{29.1 $\pm$ 1.0 mJy} at 843 MHz.  

The optical spectrum (Fig. \ref{spectra2}, upper right panel), shows a broad H$_\alpha$+[N{\sc ii}] emission complex with narrow H$_{\beta}$ emission line.
Through the H$_\beta$/[O {\sc iii}]$\lambda$5007 line flux ratio we are able to classify for the first time this source as a Seyfert 1.9 AGN with
\mbox{z = 0.123$\pm$0.001}.  
%At this redshift we derive a radio \mbox{luminosity} for this source of 1.9 $\times 10^{40}$ erg s$^{-1}$ at 1.4 GHz.
\subsection{ Swift J0739.6-3144}
Swift J0739.6-3144 is associated with the galaxy LEDA 86063, with magnitude B=16.51 and redshift 0.0257. 
It is positionally consistent with an NVSS source (NVSS 073944-314304) and a MGPS-2 source, with fluxes of 31.6 $\pm$ 1.1 mJy at 1.4 GHz
and \mbox{42.4 $\pm$ 2.2 mJy} at 843 MHz respectively.
The analysis of the optical spectrum (Fig. \ref{spectra2}, central left panel), { published here} for the first time, only shows narrow permitted and forbidden emission lines, 
{ which enable} us to classify it as a type 2 AGN with redshift 0.026$\pm$0.001,
confirming the classification reported by Ajello et al. (2008).

We now describe in detail the results from the X-ray spectral analysis of this AGN. It is found that a
simple absorbed power law fails to reproduce the spectrum of Swift J0739.6-3144 ($\chi^{2}$ = 363 for 153 d.o.f.\footnote{d.o.f. = Degrees of freedom}):
it also gives a negative value of photon index, no intrinsic absorption and an observed flux of 9.4 $\times$ 10$^{-13}$ 
erg cm$^{-2}$ s$^{-1}$ in the 2-10 keV band.
Inspection of the data residuals obtained using this simple model indicates a presence of strong soft excess below 
1 keV and a prominent line around 6.4 keV; these features, together with the extremely flat power law, 
strongly point to a highly absorbed Compton thick AGN.

Following the analysis of IGR J16351-5806, a new high-energy selected Compton thick AGN (Malizia et al. 2009),
we tried to fit the soft excess of Swift J0739.6-3144 with the \texttt{mekal} model in \texttt{XSPEC} which fits
well the soft part of the spectrum and allows 
various scenarios (transmission, reflection and complex absorption) to account for the continuum up to 10 keV.
Although this model yields an adequate fit to our spectrum,
it is worth noting that it may be an oversimplified parameterization of
the data. High resolution
spectroscopy of nearby Seyfert 2s have, in fact, demonstrated that the
soft X-ray emission is often
dominated by emission lines from photoionized gas which have a negligible contribution from an
underlying continuum. The blending of
these emission lines in the EPIC spectra can mimic a continuum emission
(Iwasawa et al. 2003).
However, as our main goal is to study the nature of Swift J0739.6-3144, the uncertainties
induced by a purely phenomenological modelling of the soft excess will
not substantially affect our results.

The addition of \texttt{mekal} to the absorbed power law improves the fit ($\Delta\chi^2$=79 for 2 d.o.f.)
and gives a gas temperature of kT of 0.65$^{+0.18}_{-0.07}$, still leaving an extremely flat power low continuum ($\Gamma$ $\sim$ 0.6).
Another substantial improvement ($\Delta\chi^2$=82 for 3 d.o.f.)  is obtained when we introduce 
the $K\alpha$ iron fluorescence emission line: the energy of the line is found to be at 
6.41 $\pm$ 0.04 keV with an equivalent width EW $<$ 1 keV. The intrinsic width of the line is 
narrow { ($\sigma \sim$ 0.15$^{+0.07}_{-0.04}$ keV)} and for simplicity it has been frozen to the observed value in subsequent fits. 
The flat slope of the power law  as well as the strength of the line at 6.4 keV 
(EW $\sim$ 0.5-1 keV) and the apparent lack of absorption again suggest that the source could be reflection dominated and therefore
highly absorbed (Matt et al. 2000). 

\begin{figure}[th!]
%\begin{center}
\hspace{-.1cm}
\centering{\mbox{\psfig{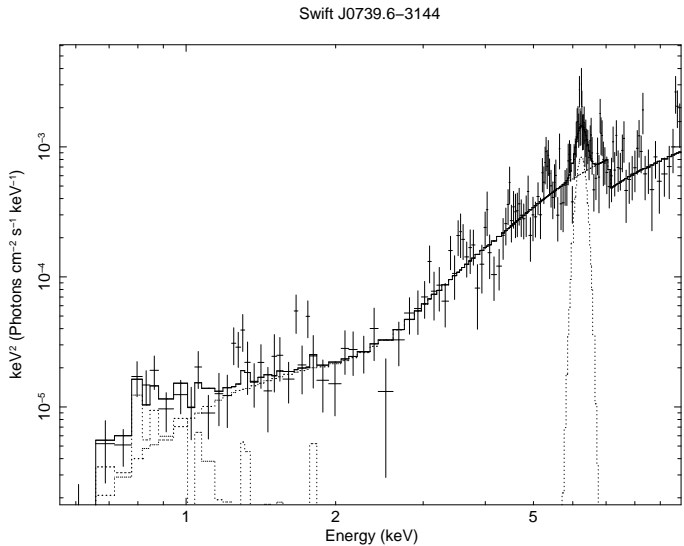}}}
%\vspace{-.3cm}
\caption{\emph{XMM} EPIC-pn X-ray spectrum of Swift J0739.6$-$3144, possible Compton thick source. The solid line shows the best fit model.
}\label{xspec}
%\end{center}
\end{figure}

The models generally used to account for the X-ray continuum of AGNs in this regime are the transmission 
model, pure reflection model and, as recently found by Malizia et al. (2009), the complex absorption model (see this { latter} reference for details).
We applied all these models to our data and at first we ruled out the pure reflection one because this model assumes that the 
absorbed power law is totally absorbed by a column density N$_{H}>$ 10$^{25}$ cm$^{-2}$ implying an EW of the $K\alpha$ iron line of 1 keV or even more,
while we still measure for our line an equivalent width { of} less than 1 keV (0.5$^{+0.12}_{-0.13}$keV).
Both the transmission and the complex absorption scenario provide a good fit to our data ($\chi^{2}$=159/148 and  $\chi^{2}$ = 159/146 respectively, this implies a chance improvement probability of $\sim 10^{-7}$ for both models according to the { F-test}, see Press et al. 1992)
but while the first still gives a quite flat photon index ($\Gamma$ = 1.25 $\pm$0.1) with a column density of about 2 $\times$  10$^{23}$ cm$^{-2}$,
the second provides a more canonical value of the power law continuum of ($\Gamma$ = 1.77$^{+0.07}_{-0.09}$). 
The absorption required by the data is in the form of two columns (N$_{H_{1}}$ $\sim$ 5 $\times$ 10$^{23}$ cm$^{-2}$ and 
N$_{H_{2}}$ $\sim$ 1.5 $\times$ 10$^{23}$ cm$^{-2}$), both covering 70-90\% of the source. 
In this case the combinations of such  columns are able to explain the observed 
iron line EW (Ghisellini et al. 1994).
In Fig. \ref{xspec} the unfolded spectrum fitted with this model is shown.

As for the case of IGR J16351-5806, also for Swift J0739.6-3144 this model is 
of interest in view of recent studies on the torus geometry and its nature, which strongly indicate that 
this structure is clumpy and made of dusty clouds (Elitzur 2008). 
Of course only the study of the broad band spectrum i.e. the combination of the X-ray data with the high energy information ($>$10 keV)
can help to definitely assess the nature of this source i.e. if it is just a highly absorbed object or a definite Compton thick AGN.

\subsection{ Swift J0743.0-2543}
This source is associated with the galaxy LEDA 86073, with magnitude B=14.0, obtained from the USNO$-$A2.0 catalog (Monet et al. 2003). 
It was detected by Ajello et al. (2008), likely associated with a ROSAT all-sky bright source (1RXS J074315.6-254545), with an IRAS source (IRAS 07411$-$2538)
and with a NVSS radio source (NVSS 074314-254547), with a flux density of 0.203 $\pm$ 0.027 mJy at 1.4 GHz.

Its optical spectrum (Fig. \ref{spectra2}, central right panel) shows a broad emission H$_\alpha$+[N{\sc ii}] complex with broad H$_\beta$ and HeI emission lines, which enable us
to classify for the first time Swift J0743.0-2543 as a Seyfert 1.2 AGN at z = 0.023$\pm$0.001.
%At this redshift, the source radio luminosity at 1.4 GHz is 1.2 $\times 10^{37}$ erg s$^{-1}$.
\subsection{ Swift J0811.5+0937}
Its optical counterpart is USNO$-$A2.0 0975-05763590 with a magnitude B=19.2 according to the USNO Catalog. It is also associated with an NVSS source (NVSS 081130+093350) with a flux density of 4.3 $\pm$ 0.5 mJy at 1.4 GHz.
The spectrum (Fig. \ref{spectra2}, bottom left panel), shows a forbidden [O{\sc ii}] narrow emission line and absorption features. 
{ Following the method of Laurent-Muehleisen et al. (1998)}, for this source we have calculated the Ca {\sc ii} break contrast at 4000 \AA~(Br$_{\rm 4000}$), as defined by Dressler \& Shectman (1987), and its value is $\sim 45\%$;  
the presence of other absorption features, such as the G band, the Mg {\sc i }and the Ca {\sc ii} H+K doublet and the lack of strong Balmer absorption lines, 
enables us to state that this source is a `normal galaxy'.
This peculiar object is thus an XBONG with redshift of 0.286$\pm$0.001. Our classification agrees with that reported in Ajello et al. (2008).
\subsection{ Swift J0902.0+6007}
The optical counterpart is LEDA 25370 with magnitude B = 14.56 and redshfit 0.0111.
This source is associated with an {\it XMM} serendipitous source (2XMMi J090158.7+600903) and is listed in the NVSS Catalog (NVSS 090158+600906)
with a flux density of 30.3 $\pm$ 1.0 mJy at 1.4 GHz. 

The optical spectrum (Fig. \ref{spectra2}, bottom right panel) shows narrow H$_\alpha$+[N{\sc ii}] emission complex, H$_{\beta}$ and H$_{\gamma}$ permitted narrow emission lines,
weak He{\sc i} and He{\sc ii} emission lines; [O{\sc iii}] and [O{\sc ii}] forbidden emission lines are also present. Through these lines, we estimate a redshift of 0.012$\pm$0.001, consistent with that reported in the Hyperleda archive.
As before, through the \mbox{diagnostic} diagrams {listed in Sect. 2} we can classify the object for the first time as a Seyfert 2 AGN.
In the Tueller et al. (2008) catalog, this source was simply classified as galaxy.
\subsection{ Swift J0904.3+5538}
The optical counterpart is USNO$-$A2.0 1425-07270668, with a magnitude R=13.5, provided in this same catalog.
This {\it Swift } source is positionally consistent with an {\it XMM} serendipitous source (2XMMi J090436.9+553602). 

Its optical spectrum showed in Fig. \ref{spe} (upper left panel) has all the typical features of a broad line AGN: a broad emission H$_\alpha$+[N{\sc ii}] complex with broad H$_\beta$ and H$_{\gamma}$ emission lines and some forbidden narrow emission lines.
Through the optical spectrum and the H$_\beta$/[O{\sc iii}]$\lambda$5007 line flux ratio, we are able to classify this source as a Seyfert 1.5 AGN at redshift 0.0374$\pm$0.0003. 
This is consistent with the redshift determination of Schneider et al. (2003) for this galaxy.
This object was reported by Winter et al. (2008) as a Seyfert 1 galaxy. 
\subsection{Swift J0911.2+4533}
The optical counterpart of this {\it Swift} source is the galaxy LEDA 2265450, with a magnitude B=16.43 and a redshift of 0.0268. 
Its radio counterpart is NVSS 091129+452804 with a flux of 3.1 $\pm$ 0.5 mJy.

The optical spectrum (Fig. \ref{spe}, upper right panel), shows a very narrow emission H$_\alpha$+[N{\sc ii}] complex, with the presence of [O{\sc iii}] and [O{\sc ii}] forbidden emission lines.
From these features we can classify this source as a Seyfert 2 AGN with a redshift of 0.0269$\pm$0.0003. 
Our classification agrees with that reported in the BAT survey of Tueller et al. (2008).  

%We can complete our classification of this source by evaluating the thickness parameter to assess its Compton nature.
%The 2--10 keV/[O{\sc iii}] flux ratio is larger than 0.8 and the softness ratio is $\sim$0.08, therefore this is a Compton thin source. 
%Its 1.4 GHz radio luminosity is 8.4 $\times 10^{37}$ erg s$^{-1}$ at the source redshift.
\subsection{ Swift J0917.2-6221}
The optical counterpart of this object is the galaxy LEDA 90443, with redshift 0.0573 and magnitude R=11.5 reported in the USNO$-$A2.0 catalog.
This source is associated with a {\it ROSAT} all-sky survey bright source (1RXS J091609.5-621934) and with the far-infrared object IRAS 09149-6206.
Moreover, this {\it Swift} object is listed in the radio catalog MGPS-2 as MGPS-2 J091609-621928 with a flux at 843 MHz of 61.4 $\pm$ 2.1mJy. 

The spectrum (see Fig. \ref{spe}, central left panel) shows very broad Balmer lines in emission, on a blue continuum, H$_\alpha$+[N{\sc ii}] complex, as well as narrow [O{\sc iii}] forbidden lines.
Through the \mbox{H$_\beta$/[O {\sc iii}]$\lambda$5007} line flux ratio, we can give a more accurate Seyfert 1.2 classification with respect to the Sy1 given by Winter et al. (2008).  
The redshift calculated from the [O {\sc iii}]$\lambda$5007 forbidden line is 0.057$\pm$0.001, in agreement with that reported in Hyperleda archive.
The source redshift implies a 843 MHz radio luminosity of  4.7 $\times 10^{39}$ erg s$^{-1}$.
\subsection{Swift J0923.7+2255}
This object is associated with the galaxy LEDA 26614, with magnitude B=15.16 and redshift 0.0323. 
It is listed in the {\it ROSAT} Bright Source Catalog (1RXS J092343.0+225437) and in the {\it XMM} Serendipitous Source Catalog (2XMMi J092342.9+225433).
It has an NVSS radio counterpart (NVSS 092343+225430), with a flux density at 1.4 GHz of 10.3 $\pm$ 0.5 mJy.
The spectrum (see Fig. \ref{spe}, central right panel) shows the H$_\alpha$+[N{\sc ii}] complex in emission, the Balmer lines, He{\sc i} permitted and some forbidden narrow lines in emission,
Indeed, Swift J0923.7+2255 has a full width at half-maximum (FWHM) of H$_{\beta}$ line of $\sim$ 2000 km s$^{-1}$, the presence of [Fe{\sc i}] emission lines and the [OIII]$_{5007}$/H$_{\beta}$ flux ratio $<$ 3.
Following the approach of Osterbrock  \& Pogge (1985) and Goodrich (1989) we are able to
classify it as a Narrow-Line Seyfert 1 AGN, at z = 0.034$\pm$0.001. 
Tueller et al. (2009) classified this source as a Seyfert 1.2 AGN.
%The radio luminosity at 1.4 GHz is 4.5 $\times 10^{38}$ erg s$^{-1}$ at the source redshift.
\subsection{ Swift J1049.4+2258}
The optical counterpart of this X-ray object is the galaxy LEDA 32398 (Mrk 417), with a B magnitude of 15.91 and redshift 0.0326. 
It is associated with a {\it XMM} 2nd serendipitous source (2XMMi J104930.9+225753).

The spectrum (Fig. \ref{spe}, lower left panel) shows a weak H$_\alpha$+[N{\sc ii}] complex, narrow permitted H$_{\beta}$ line and prominent [O{\sc iii}] forbidden emission lines, with a weak continuum.
Therefore, we classify this object as a Seyfert 2 AGN with redshift 0.033 $\pm$ 0.001, confirming what reported by Winter et al. (2008).

%Moreover, we can include some information on the thickness parameter, and therefore on its Compton nature.
%The \mbox{2--10 keV / [O {\sc iii}]} flux ratio of $\sim$ 7 and the softness ratio of $\sim$0.1 allow us to state that Swift J1049.4+2258 is a Compton thin source.
\vspace{0.3cm}

\subsection{{ Central black hole} masses}
In this section we estimate the mass of the central black hole for 6 out of 7 type 1 AGNs.
The method used here follows the prescription of Wu et al. (2004) and Kaspi et al. (2000). 
We used the H$_{\beta}$ emission flux, corrected for the Galactic color excess (Schlegel et al. 1998), and the broad-line region
(BLR) gas velocity. 

Through Eq. (2) of Wu et al. (2004) we estimate the BLR size, then we used Eq. (5) of Kaspi et al. (2000)
using the BLR size and the $v_{FWHM}$ to calculate the AGN black hole mass. The results are reported in Table \ref{blr}.
We could not estimate the mass of the central black hole of Swift J0727.5-2406 because only the narrow component of the H$_{\beta}$ line is detected for this source.
\begin{table}[h!]
\caption[]{BLR gas velocities and 
central black hole masses for 6
Seyfert 1 AGNs listed in this paper.}
\label{blr}
\begin{center}
%\vspace{-.3cm}
\begin{tabular}{lcc}
\noalign{\smallskip}
\hline
\hline
\noalign{\smallskip}
\multicolumn{1}{c}{Object} & $v_{\rm BLR}$ & $M_{\rm BH}$ \\
\multicolumn{1}{c}{}& (km s$^{-1}$)&(10$^7$ $M_\odot$)\\
\noalign{\smallskip}
\hline
\noalign{\smallskip}

Swift J0059.4+3150   & 3500 & 4.5 \\
Swift J0342.0$-$2115 &4150  & 8.3  \\
Swift J0743.0$-$2543   & 2510 & 2.6 \\
Swift J0904.3+5538& 2910  & 3.3  \\
Swift J0917.2$-$6221 & 3460 & 7.0 \\
Swift J0923.7+2255 & 1820& 6.4 \\
\noalign{\smallskip} 
\hline
\hline
\end{tabular}
\end{center}
\end{table}

\subsection{Compton thickness estimation for type 2 AGNs}
Of the 17 extragalactic objects identified in this work, 9 ($\sim$ 53\%) 
are type 2 Seyfert galaxies. 
In the {\it Swift}/BAT surveys as well as that of {\it INTEGRAL}/IBIS there are a number of 
highly absorbed Seyfert galaxies, among these there are a few Compton thick objects.
For this reason in this section we try to investigate the nature of our type 2 objects 
by using a diagnostic tool developed by Malizia et al. (2007).

\begin{figure}[th!]
%\begin{center}
\hspace{-.1cm}
\centering{\mbox{\psfig{file=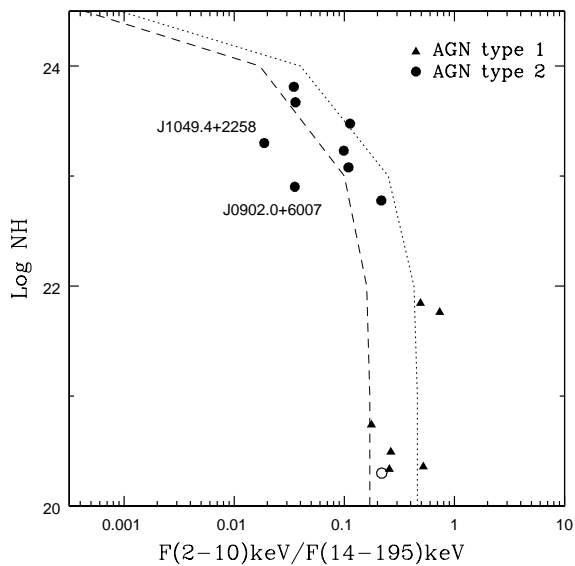,width=9cm}}}
%\vspace{-.3cm}
\caption{F$_{2-10 keV}$/ F$_{14-195 keV}$ flux ratio of our sample. Lines correspond to expected values for an absorbed power law
with photon index 1.5 (dot) and 1.9 (dash). The open circle shows the position of Swift J0501.9-3239, the Seyfert 2 without local absorption.
}\label{ratio}
%\end{center}
\end{figure}

We have performed the X-ray data analysis of all our sample sources (see Table \ref{X}),
in order to evaluate in a consistent way the absorption amount in excess to the Galactic
one and the flux in the 2-10 keV energy band.
In a similar manner to Malizia et al. (2007), plotting  the absorption 
against the softness ratio F$_{(2-10)keV}$/F$_{(14-195)keV}$  can help us to
identify Compton thick candidates (see Fig. \ref{ratio}). A clear trend of decreasing softness ratio as the absorption 
increases is visible as expected if the 2-10 keV flux is progressively depressed as the
absorption becomes stronger. In Fig. \ref{ratio} the value of N$_H$ used is that of the intrinsic absorption, when measured
(see Table \ref{X}), or the Galactic value in the cases either when no intrinsic absorption exists (mainly Seyfert 1 AGNs) or the statistics are too 
poor for it to be measured (Swift J0727.5-2406 and Swift J0811.5+0937).
%apart from those two (Swift J0727.5-2406 and Swift J0811.5+0937) 
%for which the intrinsic N$_H$ is not available, have been plotted and the type 2 have been distinguished from type 1, 
Lines represent the expected values for an absorbed power law with photon index of 1.5 (dotted line) and
1.9 (dashed line). 

Most of our sources follow the expected trend, 
none of them turns out to have a low softness ratio with respect to the 
estimated column density; only two (Swift J0902.0+6007 and Swift J1049.4+2258)
are slightly outside the expected trend, but probably in both cases the low quality statistics of
the XRT data do not allow us to properly measure their column densities.
It is important to note that, as expected the type 1 objects occupy the low part of the diagram (triangles in the figure),
while the majority of our type 2 objects (circles) turn out to be highly absorbed ($>$ 10$^{23}$ cm$^{-2}$) AGNs; none of which appears to
be a Compton thick object. 

As reported in the Appendix a good quality XMM-\emph{Newton} spectral measurement was available only for Swift J0739.6$-$3144,
thus enabling its nature to be defined. For the rest of the sources the quality of the X-ray data is quite low and future deep X-ray observations, possibly together with
high energy ($>$10 keV) data, are needed to assess their nature. 
Finally, Swift J0501.9$-$3239 (open circle) is intriguing, since as reported in section 3.6 we find this source to be a Seyfert 2,
but from our XRT data analysis no absorption intrinsic to the source has been measured.
Its softness ratio ($\sim$ 0.26) is not particularly low, therefore it is unlikely to be a Compton thick AGN.
Also in this case higher X-ray data quality are needed to assess if it is a real unabsorbed Seyfert 2 galaxy.

\section{Conclusions}
In this work we have either given for the first time, or confirmed, or corrected, the optical spectroscopic identification of 17 {\it Swift} AGNs. 
This was achieved through a multisite observational campaign in Europe, Central and South America.

We found that our sample is composed of 16 AGNs (7 of Type 1 and 9 of Type 2) and 1 XBONG, with redshifts between 0.012 and 0.286.
For all our sources the X-ray data analysis has been performed in order to evaluate their main spectral parameters ($\Gamma$, $N_{H}$
and 2-10 keV flux).
The measurements of the column densities and the soft X-ray fluxes, together with the hard X-ray ones provided by
Swift-BAT, allowed us to use the diagnostic tool developed by Malizia et al. (2007) in order to pinpoint Compton thick candidates.
Our Seyfert 2 objects turn out to be highly absorbed (N$_H\sim 10^{23}$ cm$^{-2}$),{but none of them are Compton thick.}

%We can state that 16 of them are Compton thin AGNs and 1 is very likely a Compton thick source. Of the 16 AGNs, one (Swift J0501.9-3239) is a naked Seyfert 2 galaxy. 
%If we consider our entire sample, we see that the 94$\%$ of them are Compton thin and only the 6$\%$ are Compton thick.
%This group of sources is constituted by only 17 objects, selected with criteria which do not allow us to consider it as a complete sample an which is also affected by non-negligible statistical errors, given that it is relatively small. 
%Nevertheless, this result is in agreement with those obtained from the statistical analysis of extragalactic hard X-ray sources such as those of Markwardt et al. (2005), Beckmann et al. (2006),
%Bassani et al. (2006), Sazonov et al. (2007), Ajello et al. (2008), Tueller et al. (2008) and Paltani et al. (2008), who find an occurrence of 5-10$\%$ of Compton thick objects in hard X-ray survey performed with {\it INTEGRAL} or {\it Swift}.
Moreover, for six type 1 AGNs we have estimated the BLR size, velocity and the central black hole mass.

All of the results shown in this work stress the importance of the optical spectroscopic followup, not only for the classification of unidentified sources and for the study of the statistical properties of the various source classes,
but also for the search of Compton thick AGNs, that are thought to provide an important contribution to the overall cosmic energy budget at hard X-rays, but for which the cosmological evolution and space density are not well known yet. 
\begin{acknowledgements}

We thank Silvia Galleti for Service Mode observations at the Loiano 
telescope; Francesca Ghinassi for service mode observations 
at the TNG; Hripsime Navasardyan for service mode observations at the
Asiago Telescope; Antonio De Blasi and Ivan Bruni for night assistance at the 
Loiano telescope; Edgardo Cosgrove, Manuel Hern\'andez and Jos\'e 
Vel\'asquez for day and night assistance at the CTIO telescope.
We also thank the anonymous referee for useful comments which helped us to improve the quality of this paper.
This research has made use of the ASI Science Data Center Multimission 
Archive, of the NASA Astrophysics Data System Abstract Service, 
the NASA/IPAC Extragalactic Database (NED), of the NASA/IPAC Infrared 
Science Archive, which are operated by the Jet Propulsion Laboratory, 
California Institute of Technology, under contract with the National 
Aeronautics and Space Administration and of data obtained from the High Energy 
Astrophysics Science Archive Research Center (HEASARC), provided by NASA's GSFC.
This publication made use of data products from the Two Micron All 
Sky Survey (2MASS), which is a joint project of the University of 
Massachusetts and the Infrared Processing and Analysis Center/California 
Institute of Technology, funded by the National Aeronautics and Space 
Administration and the National Science Foundation.
This research has also made use of data extracted from the 6dF 
Galaxy Survey and the Sloan Digitized Sky Survey archives;
the SIMBAD database operated at CDS, Strasbourg, 
France, and of the HyperLeda catalogue operated at the Observatoire de 
Lyon, France.
The authors acknowledge the ASI and INAF financial support via grants No. I/023/05/0 and I/008/07;
P.P. is supported by the ASI-INTEGRAL grant No. I/008/07.
L.M. is supported by the University of Padua through grant No. 
CPDR061795/06. G.G. is supported by FONDECYT 1085267.
V.C. is supported by the CONACYT research grant 54480-F (M\'exico).
D.M. is supported by the Basal CATA PFB 06/09, and FONDAP Center for 
Astrophysics grant No. 15010003.

\end{acknowledgements}

\end{document}